\documentclass[a4paper,reqno,12pt,final]{amsart}
\usepackage{amsmath,amsthm,amssymb,amscd,euscript,bbold}
\usepackage{array}
\usepackage[T1]{fontenc}
\usepackage[latin1]{inputenc}
\usepackage[notref,notcite,color]{showkeys}
\newcommand{\RR}{\mathbb{R}}
\newcommand{\CC}{\mathbb{C}}

\newcommand{\OO}{\mathbb{O}}
\newcommand{\Spin}{\mathrm{Spin}}

\newcommand{\SL}{\mathrm{SL}}
\newcommand{\SO}{\mathrm{SO}}
\newcommand{\ISO}{\mathrm{ISO}}
\newcommand{\SU}{\mathrm{SU}}
\newcommand{\so}{\mathfrak{so}}
\renewcommand{\Sp}{\mathrm{Sp}}
\DeclareMathOperator{\Cl}{C\ell}
\DeclareMathOperator{\Mat}{Mat}
\newcommand{\M}{\mathsf{M}}
\newcommand{\MM}{\mathbb{M}}
\newcommand{\VV}{\mathbb{V}}
\newcommand{\1}{\mathbb{1}}
\newcommand{\fn}{\mathfrak{n}}
\newcommand{\fg}{\mathfrak{g}}
\newcommand{\fk}{\mathfrak{k}}
\newcommand{\fh}{\mathfrak{h}}
\newcommand{\btheta}{\boldsymbol{\theta}}
\newcommand{\bomega}{\boldsymbol{\omega}}
\newcommand{\bOmega}{\boldsymbol{\Omega}}
\newcommand{\half}{\tfrac{1}{2}}
\newcommand{\repre}[1]{\underline{\mathbf{#1}}}
\newcommand{\vol}{\mathrm{vol}}
\newcommand{\sssection}[1]{\subsubsection{\underline{#1}}}
\renewcommand{\d}{\partial}

\DeclareMathOperator{\tr}{tr}
\theoremstyle{plain}

\theoremstyle{definition}

\theoremstyle{remark}

\begin{document}

\title{Breaking the $\M$-waves}
\author[JM Figueroa-O'Farrill]{José Miguel Figueroa-O'Farrill}
\address{\begin{flushright}Department of Physics\\
Queen Mary and Westfield College\\
London E1 4NS, UK\end{flushright}}
\email{j.m.figueroa@qmw.ac.uk}
\date{\today}
\begin{abstract}
  We present a systematic attempt at classification of supersymmetric
  $\M$-theory vacua with zero flux; that is, eleven-dimensional
  lorentzian manifolds with vanishing Ricci curvature and admitting
  covariantly constant spinors.  We show that there are two distinct
  classes of solutions: static spacetimes generalising the
  Kaluza--Klein monopole, and non-static spacetimes generalising the
  supersymmetric wave.  The classification can be further refined by
  the holonomy group of the spacetime.  The static solutions are
  organised according to the holonomy group of the spacelike
  hypersurface, whereas the non-static solutions are similarly
  organised by the (lorentzian) holonomy group of the spacetime.
  These are subgroups of the Lorentz group which act reducibly yet
  indecomposably on Minkowski spacetime.  We present novel
  constructions of non-static vacua consisting of warped products of
  $d$-dimensional $pp$-waves with ($11{-}d$)-dimensional manifolds
  admitting covariantly constant spinors.  Our construction yields
  local metrics with a variety of exotic lorentzian holonomy groups.
  In the process, we write down the most general local metric in
  $d\leq 5$ dimensions describing a $pp$-wave admitting a covariantly
  constant spinor.  Finally, we also discuss a particular class of
  supersymmetric vacua with nonzero four-form obtained from the
  previous ones without modifying the holonomy of the metric.  This is 
  possible because in a lorentzian spacetime a metric which admits
  parallel spinors is not necessarily Ricci-flat, hence supersymmetric 
  backgrounds need not satisfy the equations of motion.
\end{abstract}
\maketitle

\tableofcontents

\section{Introduction}

An important aspect of modern string theory is the study of its
supersymmetric solitons.  In the appropriate limit this involves the
study of solutions to the equations of motion of the relevant
supergravity theory, and in particular of those solutions which are
left invariant by some of the supersymmetries.  By now a large class
of solutions are known, including waves, monopoles and branes; but as
no systematic study has yet been undertaken, it is hard to estimate
how much of the space of all supersymmetric solutions do the known
ones comprise.

This seems to us an interesting question to address and in this paper
we start such a systematic analysis, focusing as a first step on
purely gravitational supersymmetric solutions; although later in the
paper we will also discuss vacua with nonzero flux.  As we will review
briefly below supersymmetric bosonic vacua with zero flux are given by
Ricci-flat lorentzian manifolds admitting covariantly constant
spinors.  We will see that there are two classes of solutions,
depending on whether or not the spacetime admits a covariantly
constant time-like vector.  Those spacetimes which do are static and
can be understood as generalisations of the Kaluza--Klein monopole.
Their classification reduces (up to discrete quotients) to the
classification of $10$-dimensional riemannian manifolds with holonomy
contained in $\SU(5)$, i.e., to the classification of Calabi--Yau
$5$-folds.  The second class of solutions consists of spacetimes which
are not static, but which nevertheless have a covariantly constant
light-like vector.  They can be understood as supersymmetric
gravitational $pp$-waves.  This second type of solutions will be the
main focus of this paper.  We will present a new class of
supersymmetric waves.  The simplest examples preserve half of the
supersymmetry and are built as warped products of gravitational plane
waves and flat euclidean space.  Replacing euclidean space by any
manifold admitting covariantly constant spinors, one obtains other
solutions which preserve a smaller fraction of the supersymmetry.
These solutions are purely gravitational and possess null isometries.
As we will see, they owe their existence to some exotic holonomy
groups which lorentzian spacetimes can possess.

For definiteness, we will only consider eleven-dimensional
supergravity, the low-energy limit of $\M$-theory, by which we mean
the strong coupling limit of type IIA superstring theory.  This is a
relatively simple supergravity theory which yields many of the lower
dimensional supergravity theories after dimensional reduction.  Many
of the supersymmetric solutions of the other supergravity theories can
be obtained from those of eleven-dimensional supergravity via a
mixture of dimensional reduction, dualities and dimensional oxidation.
Therefore it represents only a minor loss of generality to focus our
attention on this theory, and in fact much of what we will say will
generalise to other supergravity theories.

In summary, the purpose of this paper is then to study a class of
spacetimes corresponding to gravitational waves possessing a
covariantly constant spinor.  These spacetimes possess a covariantly
constant null vector, hence they are special cases of the spacetimes
discussed by Brinkmann in the 1920s.  We will focus primarily on a
class of solutions which can be understood as lifts to eleven
dimensions of lower dimensional $pp$-waves admitting parallel spinors.
We will write down the most general local metric for such $pp$-waves
in $d\leq 5$ dimensions, and we will investigate their lifts to
supersymmetric vacua of $\M$-theory preserving a fraction of the
supersymmetry that can be easily computed.  As an illustration of the
construction we will work out one example in detail, corresponding to
the lift to eleven dimensions of the Nappi--Witten plane wave
\cite{NW}.

This paper is organised as follows.  In Section~\ref{sec:setup} we
briefly review eleven-di\-men\-sion\-al supergravity and recast the
problem of finding supersymmetric vacua in geometrical terms.  The
possible vacua can be organised according to their holonomy group and
we give two tables of such holonomy groups with the fraction of
supersymmetry preserved in each case.  Some of the holonomy groups
appearing in the tables may not be familiar, so we also briefly
discuss lorentzian holonomy groups at the end of that section.  Also
in an appendix to the paper we discuss in detail the relevant
lorentzian holonomy groups.  In Section~\ref{sec:pp} we discuss some
relevant facts about $pp$-waves, i.e., spacetimes admitting null
parallel vectors, and in particular about supersymmetric $pp$-waves:
the subclass also admitting parallel spinors.  We write down the most
general local metric describing a supersymmetric $pp$-wave in
dimension $d\leq 5$.  We also review Bryant's recent construction of
the most general local metric in eleven dimensions which admits a
parallel null spinor (see below).  Bryant's construction in principle
solves the problem of constructing the most general supersymmetric
$\M$-theory vacuum with vanishing four-form; but since the result is
not constructive, it is useful to have explicit constructions of such
vacua.  In Section~\ref{sec:lift} we discuss a method of constructing
such vacua, consisting in the lift to eleven dimensions of a
lower-dimensional $pp$-wave admitting parallel spinors.  We discuss
the lifting of such a $d$-dimensional $pp$-wave to eleven dimensions.
Examples of Bryant's metrics possessing (conjecturally) all possible
holonomy groups of $\M$-waves are given by warping a three-dimensional
supersymmetric $pp$-wave with an appropriate eight-dimensional
manifold.  In Section~\ref{sec:NW} we discuss one such example in
detail: the four-dimensional Nappi--Witten plane wave.  We discuss its
geometry, compute its holonomy and prove that it admits parallel
spinors, despite having non-vanishing Ricci curvature.  We then lift
the Nappi--Witten geometry to eleven dimensions by warping it with an
appropriate seven-dimensional manifold.  The resulting metric solves
the supergravity equations of motion and preserves a fraction of the
supersymmetry which can be as much as $\half$ and as little as
$\tfrac1{16}$, depending on the holonomy of the seven-manifold.  In
Section~\ref{sec:addflux} we investigate the possibility of adding a
nonzero four-form to the metrics constructed above while still
satisfying the equations of motion and preserving some supersymmetry.
We will see this is possible for most of the metrics constructed
above.  Finally in Section~\ref{sec:final} we offer some concluding
remarks and point out some open questions.

\section*{Note added in proof}

After this preprint had been circulated and submitted to the archive,
we became aware of the paper \cite{Lidsey} by Lidsey, who also
considers building eleven-dimensional Ricci-flat spacetimes by
embedding a wide class of lower-dimensional $pp$-waves.

\section{Statement of the problem and holonomy analysis}
\label{sec:setup}

In this section we set up the problem.  In order to do so we start
with a brief review of eleven-dimensional supergravity.  We then
recast the search for purely gravitational bosonic vacua as the
problem of constructing Ricci-flat metrics with prescribed holonomy.
We discuss the holonomy classification of supersymmetric vacua.

\subsection{A brief review of eleven-dimensional supergravity}
\label{sec:d=11sugra}

The arena of eleven-dimensional supergravity \cite{Nahm,CJS} is a
manifold $M$ with a metric $g$ of signature $10{+}1$ and a spin
structure.  From now on, we will call such a manifold simply a
lorentzian spin manifold, leaving its dimension implicitly equal to
eleven, unless otherwise stated.  Apart from the metric, the other
fields in the theory are a closed $4$-form $F$, which is given locally
in terms of a $3$-form potential $A$ by $F=dA$, and the gravitino
$\psi$.  In this paper we will only be interested in bosonic vacua,
for which the gravitino vanishes: $\psi=0$; although some comments
will be made in Section~\ref{sec:addgrav} about the general case.  The
bosonic part of the action is given by a sum of three terms: an
Einstein--Hilbert term, a generalised Maxwell term and a Chern--Simons
term, as follows:
\begin{equation}
  \label{eq:action}
  I = \half \int d^{11}x \sqrt{-g} R  - \tfrac14 \int F\wedge\star F -
  \tfrac1{12} \int A \wedge F \wedge F~.
\end{equation}
We could also add a gravitational Chern--Simons term
\begin{equation}
  \label{eq:gravCS}
  \half \int A \wedge X_8~,
\end{equation}
where the closed 8-form $X_8$ is given by
\begin{equation}
  \label{eq:X8}
  X_8 = \tfrac1{192} \tr R^4 - \tfrac1{768} \left(\tr R^2\right)^2~;
\end{equation}
but we will ignore this term in what follows, as it is a higher order
correction.

The generalised Maxwell equations following from the action are
\begin{equation}
  \label{eq:feom}
  d\star F = -\half F\wedge F~,
\end{equation}
which together with the Bianchi identity $dF=0$ specify $F$.  The
equation of motion for the metric is the generalised Einstein--Maxwell 
equation:
\begin{equation}
  \label{eq:genEM}
  R_{ab} - \half R\, g_{ab} = \tfrac16 T_{ab}(F)~,
\end{equation}
where the energy-momentum tensor of the Maxwell field is given by
\begin{equation}
  \label{eq:emt}
  T_{ab}(F) = F_{acde} F_b{}^{cde} - \tfrac18 g_{ab} F_{cdef}
  F^{cdef}~.
\end{equation}
(Notice that in eleven dimensions it is not traceless.  In fact it is
traceless only in eight dimensions, where there can be dyonic objects
contributing to $T_{ab}$.)

With the gravitino set to zero, the supersymmetry variations of the
bosonic fields vanish automatically.  This means that the conditions
for preservation of supersymmetry is simply that the supersymmetry
variation of the gravitino should remain zero.  This is equivalent to
the existence of spinors $\varepsilon$ which are parallel (i.e.,
covariantly constant) with respect to a generalised connection
\begin{equation}
  \label{eq:gravshift}
  \Hat\nabla_a \varepsilon \equiv \nabla_a \varepsilon - \theta_a(F)
  \cdot \varepsilon = 0~,
\end{equation}
where\footnote{Throughout the paper, we shall adorn ``flat'' indices
  with a $\Hat{\phantom{a}}$.} $\nabla_a = \d_a - \frac14
\omega_{a}{}^{\Hat a\Hat b}\Gamma_{\Hat a \Hat b}$ is the spin
connection, and $\theta_a(F)$ is the $F$-dependent part of the
connection:
\begin{equation}
  \label{eq:cliffconn}
  \theta_a(F) \equiv \tfrac1{288} F_{bcde} \left( {\Gamma_a}^{bcde} - 8
  \delta_a{}^b \Gamma^{cde} \right)~.
\end{equation}

The connection $\Hat\nabla$ is a very different object from the spin
connection $\nabla$.  As its name indicates, $\nabla$ only depends on
how $\varepsilon$ transforms under the spin group.  This is evident
from the fact that its $\Gamma$ matrix dependence is only through
$\Gamma_{\Hat a \Hat b}$, which are the infinitesimal generators of
$\Spin(10,1)$.  In contrast, $\Hat\nabla$ depends on terms containing
antisymmetrised products of three and five $\Gamma$ matrices.
Therefore, whereas the connection $\nabla$ takes values in the spin
subalgebra of the Clifford algebra $\Cl(10,1)$, the connection
$\Hat\nabla$ takes values in the Clifford algebra itself.  We
therefore refer to it as a \emph{Clifford connection}.  The Clifford
connection, unlike the spin connection, is not related to a connection
on the tangent bundle, which makes its analysis much more complicated.

As a first step we will therefore set $F=0$ in this paper and study
the possible supersymmetric bosonic solutions to the resulting
equations of motion.  Later in Section~\ref{sec:addflux} we will relax
this condition and investigate whether it is possible to add $F$ to
solutions already found.  From equation \eqref{eq:genEM}, it follows
that bosonic solutions with $F=0$ correspond to lorentzian spin
manifolds with vanishing Ricci curvature: $R_{ab}=0$.  From equation
\eqref{eq:gravshift} such a solution will preserve supersymmetry if in
addition the spacetime admits parallel spinors relative to the spin
connection.  In other words, in geometrical terms, we have that purely
gravitational supersymmetric solutions of eleven-dimensional
supergravity are in one-to-one correspondence with eleven-dimensional
lorentzian spin manifolds admitting parallel spinors.

Unlike the riemannian case, in a lorentzian spacetime Ricci-flatness
is \emph{not} an integrability condition for the existence of parallel
spinors.  Indeed suppose that $\varepsilon$ is a nonzero parallel
spinor: $\nabla_a \varepsilon = 0$.  Iterating this equation we find
the following integrability condition:
\begin{equation}
  \label{eq:holalg}
  R_{ab}{}^{cd} \Gamma_{cd} \varepsilon = 0~.
\end{equation}
Contracting with $\Gamma^b$ and discarding the $\Gamma$-trilinear
terms by virtue of the first Bianchi identity ($R_{[abc]}{}^d = 0$),
we obtain
\begin{equation}
  \label{eq:integrability}
  R_{ab}\Gamma^b \varepsilon = 0~.
\end{equation}
Multiplying this equation with $R_{ac}\Gamma^c$, but not summing on
$a$, we obtain
\begin{equation}
  \label{eq:riccinull}
  R_{ab} R_{ac} g^{bc} \varepsilon = 0~,
\end{equation}
which implies that, for each $a$, the vector field with components
$R_a{}^b$ is null.  We will say that such a manifold is
\emph{Ricci-null}.  Since in a riemannian manifold there are no null
vectors, this implies that $R_{ab}=0$, i.e., riemannian Ricci-null
manifolds are Ricci-flat.  In contrast, there are Ricci-null
lorentzian manifolds which are not Ricci-flat.  We will see below many
examples of such manifolds.  Notice that contracting equation
\eqref{eq:integrability} with $\Gamma^a$ shows that the Ricci scalar
does vanish.

In summary, in searching for supersymmetric bosonic vacua, it will not
be enough to check for the existence of parallel spinors, but one must
impose the equations of motion separately.  This should lay to rest
the widespread folklore that supersymmetric backgrounds automatically
satisfy the equations of motion.

\subsection{Holonomy classification of known solutions}
\label{sec:holonomy}

As we have just seen, any Ricci-flat manifold admitting a parallel
spinor is a supersymmetric vacuum of $\M$-theory with zero flux, and
all such vacua are of that form.  The existence of parallel spinors
imposes strong restrictions on the geometry and we start off by
investigating which kinds of eleven-dimensional geometries admit
parallel spinors.  The proper tool to analyse the constraints imposed
by the existence of parallel spinors is the holonomy group.  In this
section we will provide a preliminary holonomy analysis of the
problem.  As we will see, there are two types of solutions,
generalising the Kaluza--Klein monopole and the $pp$ wave,
respectively.

The relevance of riemannian holonomy groups in studying supergravity
vacua is of course well-established (see, e.g., \cite{DNP}).  In this
paper it is however the less studied lorentzian holonomy groups which
play a fundamental role.  These groups are more exotic and have not
been discussed much in the literature; although see \cite{HoTs} for
some comments on the relevance of these groups in a context not
unrelated to the present one.

In general, the existence of parallel tensors or spinors in a manifold
imposes restrictions on the holonomy group of the manifold.  In a
(pseudo)riemannian spin manifold of signature $s{+}t$, we can express
tensors and spinors relative to pseudo-orthonormal frames
$\{\boldsymbol{e}_{\Hat a}\}$, and in effect set up a correspondence
between these geometrical objects and representations of the Lorentz
group $\SO(s,t)$ or more precisely of its spin cover.  Let $T$ be a
tensor or spinor field transforming according to some representation
$\varrho$ of the spin group, and suppose that $T$ is parallel with
respect to the Levi-Cività connection: $\nabla_a T = 0$.  The
integrability condition for this equation is that the curvature
satisfies
\begin{equation*}
  R_{ab}{}^{\Hat c \Hat d} \Lambda_{\Hat c \Hat d} \cdot T =
  0\quad\text{for all $a,b$,}
\end{equation*}
where $\Lambda_{\Hat c \Hat d}$ are the generators of the Lie algebra
$\so(s,t)$ in the representation $\varrho$ under which $T$ transforms.
This means that the subalgebra $\fh \subset \so(s,t)$ generated by the
$R_{ab}{}^{\Hat c\Hat d}$ for all $a,b$ has a singlet in the
representation $\varrho$.  The Ambrose--Singer theorem says that the
Lie algebra of the holonomy group of the manifold is isomorphic to
$\fh$; thus the existence of a parallel tensor constraints the
holonomy algebra.  There is a local converse to this.  The fact that
the holonomy algebra $\fh$ has a singlet in a representation $\varrho$
guarantees the existence of a local parallel tensor $T$ transforming
according to $\varrho$.  If the manifold is simply-connected then all
obstructions to integrating the equation $\nabla_a T= 0$ vanish, and
the parallel tensor $T$ exists globally.

With these prefatory remarks behind us, let us now investigate how the
holonomy group of the spacetime is constrained by the existence of a
parallel spinor.  Let $M$ be an eleven-dimensional lorentzian spin
manifold admitting a parallel spinor $\varepsilon$.  The integrability
condition \eqref{eq:holalg} constraints the holonomy group of the spin
connection to be (conjugate to) a subgroup $H\subset \Spin(10,1)$
leaving the spinor $\varepsilon$ invariant.  Therefore we are
interested in answering the following question:
\begin{quotation}
\emph{Which subgroups of $\Spin(10,1)$ leave a spinor invariant?}
\end{quotation}
Bryant \cite{Bryant-spinors} (see also \cite{AFSS}) has answered this
question: there are two (conjugacy classes of) maximal subgroups of
$\Spin(10,1)$ which leave a spinor invariant.  Let $\varepsilon$ be a
spinor of $\Spin(10,1)$ and consider the vector $v$ with components
$v^a = \bar\varepsilon \Gamma^a \varepsilon$.  Since $\varepsilon$ is
$H$-invariant, so is $v$; whence $H$ is contained in the subgroup of
the spin group which leaves $v$ invariant: the ``little group'' of
$v$.  Just as in the more familiar case of four dimensions, these
subgroups can be distinguished by the type of the vector $v$:
space-like, time-like or null (i.e., light-like).  In fact, only two
of the three possibilities occur.  One can show that the Minkowski
norm $v^2 \equiv v^a v_a$ of this vector is negative semi-definite:
$v^2 \leq 0$, which means that either $v$ is time-like, so that $v^2
<0$, or $v$ is null, so that $v^2=0$.  This dichotomy gives rise to
the two types of supersymmetric vacua we will study in this paper.

\subsubsection{Static vacua}

If $v$ is time-like then $H$ must be contained in the subgroup
$\Spin(10) \subset \Spin(10,1)$ leaving a time-like vector invariant.
In fact, it is not hard to show that $\varepsilon$ is left invariant
by an $\SU(5)$ subgroup of $\Spin(10,1)$.  Spacetimes with holonomy
groups $H\subset \SU(5)$ are automatically Ricci-flat and hence
satisfy the supergravity equations of motion.  Such spacetimes contain
a time-like Killing vector and hence are stationary.  Moreover because
the Killing vector is actually parallel, it is hypersurface orthogonal
so that the spacetime is static.  It is not hard to show that such
spacetimes are locally isometric to a product $\RR \times X$ with
metric
\begin{equation*}
  ds^2 = - dt^2 + ds^2(X)~,
\end{equation*}
where $X$ is any riemannian $10$-manifold with holonomy contained in
$\SU(5)$; that is, a Calabi--Yau $5$-fold.  The amount of
supersymmetry which such a spacetime will preserve will depend on the
number of parallel spinors.  Assuming for simplicity that the manifold
$X$ is simply connected, one has a number of possibilities which are
summarised in Table~\ref{tab:static}.  The notation in the table is as
follows.  Assuming that $M$ is simply connected (otherwise this
applies to its universal cover), it is given by a product
\begin{equation}
  \label{eq:static}
  M = \MM^{11-d} \times K_d~,
\end{equation}
where $\MM^{11-d}$ is ($11-d$)-dimensional Minkowski spacetime.  The
table then lists the dimension $d$ of $K$, the holonomy group $H
\subset \Spin(d)$ of $K$ (and hence of $M$) and the fraction $\nu$ of
the supersymmetry that such a geometry preserves.  The fraction $\nu$
is related to the dimension $N$ of the space of $K$-singlets in the
spinor representation of $\Spin(10,1)$ by $N = 32\nu$.

\begin{table}[h!]
\centering
\renewcommand{\arraystretch}{1.3}
\begin{tabular}{|>{$}r<{$}|>{$}c<{$}|>{$}c<{$}|}\hline
  d & H \subset \Spin(d) & \nu\\
  \hline\hline
  10 & \SU(5) & \frac1{16}\\
  10 & \SU(2) \times \SU(3) & \frac18\\[3pt]
   8 & \Spin(7) & \frac1{16}\\
   8 & \SU(4) & \frac18\\
   8 & \Sp(2) & \frac 3{16}\\
   8 & \Sp(1) \times \Sp(1) & \frac14\\[3pt]
   7 & G_2 & \frac18\\[3pt]
   6 & \SU(3) & \frac14\\[3pt]
   4 & \SU(2) \cong \Sp(1) & \half\\[3pt]
   0 & \{1\} & 1\\ \hline
\end{tabular}
\vspace{8pt}
\caption{Static $\M$-theory vacua with $F=0$.  The geometry is of the
  form given by equation \eqref{eq:static} where $d$, the holonomy $H$
  of $K$ and  the supersymmetry fraction $\nu$ are listed above.}
\label{tab:static}
\end{table}

The maximal supersymmetric vacuum ($\nu=1$) corresponds to flat space,
and the half-BPS states ($\nu=\half$) corresponds to the Kaluza--Klein
monopole and its generalisations, where the spacetime is of the form
$\MM^7 \times K$, with $K$ a hyperkähler $4$-manifold.  The vacua in
the table are of course interesting, but we will not consider them
further in the present paper, since their classification is reduced,
at least locally, to the classification of Calabi--Yau $5$-folds, and
we have nothing new to add to that problem here.

\subsubsection{Non-static vacua}

On the other hand, suppose that $v$ is null.  Following Bryant let us
call such $\varepsilon$ a \emph{null spinor}.  The isotropy subgroup
(i.e., the ``little group'') of a null spinor is contained in the
isotropy subgroup of the null vector $v$, which in eleven dimensions
is isomorphic to the spin cover of $\ISO(9) = \SO(9) \ltimes \RR^9
\subset \SO(10,1)$.  Indeed it is shown in the appendix that
$\varepsilon$ is left invariant by a somewhat exotic $30$-dimensional
non-reductive Lie group isomorphic to
\begin{equation}
  \label{eq:exotic}
  G = \left(\Spin(7) \ltimes \RR^8\right) \times \RR \subset
  \Spin(10,1)~,
\end{equation}
where $\Spin(7)$ acts on $\RR^8$ according to the spinor
representation.  (See the appendix for the details about this group.)
It is not immediately obvious that $G$ is a possible holonomy group
for an eleven-dimensional spacetime, but Bryant \cite{Bryant-spinors}
has recently proven that this is the case and has moreover written
down the most general local metric with this holonomy (see equation
\eqref{eq:bryant} below).  In any case we will construct plenty of
examples in this paper.  Spacetimes with holonomy in $G$ are not
automatically Ricci-flat, but those which are will be supersymmetric
vacua of $\M$-theory.  If the holonomy is precisely $G$, then such
vacua will preserve $\tfrac1{32}$ of the supersymmetry, but one can
preserve more supersymmetry by restricting the holonomy to smaller
subgroups of $G$.  In sharp contrast to the case of static vacua
discussed above, restricting the holonomy will not necessarily
decompose the spacetime into a metric product.  In fact we will see
below several new examples of half-BPS vacua with indecomposable
metrics.

\begin{table}[h!]
\centering
\renewcommand{\arraystretch}{1.3}
\begin{tabular}{|>{$}c<{$}|>{$}c<{$}|}\hline
  H \subset \Spin(10,1) & \nu\\
  \hline\hline
  \left(\Spin(7) \ltimes \RR^8\right) \times \RR & \frac1{32}\\
  \left(\SU(4) \ltimes \RR^8\right) \times \RR & \frac1{16}\\
  \left(\Sp(2) \ltimes \RR^8\right) \times \RR & \frac3{32}\\
  \left(\Sp(1) \ltimes \RR^4\right) \times \left(\Sp(1) \ltimes
    \RR^4\right) \times \RR & \frac18\\[3pt]
  \left(G_2 \ltimes \RR^7\right) \times \RR^2 & \frac1{16}\\[3pt]
  \left(\SU(3) \ltimes \RR^6 \right) \times \RR^3 & \frac18\\[3pt]
  \left(\Sp(1) \ltimes \RR^4\right) \times \RR^5 & \frac14\\[3pt]
  \RR^9 & \half \\[3pt]
 \hline
\end{tabular}
\vspace{8pt}
\caption{Indecomposable $\M$-theory vacua with $F=0$ in terms of their 
  holonomy group $H$ and the fraction $\nu$ of supersymmetry
  preserved.}
\label{tab:mwaves}
\end{table}

Table~\ref{tab:mwaves} lists holonomy groups of simply-connected
indecomposable metrics in eleven dimensions admitting parallel spinors,
together with the fraction $\nu$ of the supersymmetry that such
spacetimes preserve.  The fraction $\nu$ is related to the dimension
$N$ of the space of $H$-invariant spinors by $N = 32\nu$.  The
holonomy groups in the table are all of the form
\begin{equation*}
  \left( K \ltimes \RR^d\right) \times \RR^{9-d}~,
\end{equation*}
where $K \subset \Spin(d)$ acts irreducibly on $\RR^d$, except for $K
= \Sp(1) \times \Sp(1) \subset \Spin(8)$ which acts reducibly on
$\RR^8$, so that one has
\begin{equation*}
  \left( \Sp(1) \times \Sp(1) \right) \ltimes \RR^8 \cong \left(\Sp(1)
    \ltimes \RR^4\right) \times \left(\Sp(1) \ltimes \RR^4\right)~.
\end{equation*}
Although we believe this table to be complete, we will not attempt a
proof here.  The reason why this is not straight-forward to check is
that the possible holonomy groups of lorentzian manifolds are not
classified.  The rest of this paper is devoted to a construction of
metrics with the holonomy groups in this table, and in particular to
those which are Ricci-flat; but before doing so we will put these
results in their proper mathematical context, by discussing lorentzian
holonomy groups.

\subsection{Lorentzian holonomy groups}

In contrast to the case of riemannian holonomy, of which a fairly
complete picture has emerged in the last half of the century, the
situation with lorentzian holonomy groups is very different.  See
\cite{BBI} for a recent survey of known results.  In this section we
would like to explain why this is the case.  This will also serve to
put some of our results in their proper mathematical context.

Let $M$ be a connected and simply-connected manifold with a
pseudo-riemannian metric of signature $s{+}t$.  The holonomy group of
$M$ is (conjugate to) a subgroup of $\SO(s,t)$.  A natural question is 
then:
\begin{quotation}
\emph{Which subgroups $H \subset \SO(s,t)$ can appear as holonomy
groups?}
\end{quotation}
Suppose that $M = M_1 \times M_2$ is isometric to a product of two
pseudo-riemannian manifolds of signatures $s_1{+}t_1$ and $s_2{+}t_2$
respectively, with $s= s_1 + s_2$ and $t=t_1 + t_2$.  Then the
holonomy group $H$ of $M$ breaks up as $H = H_1 \times H_2$, where
$H_1 \subset \SO(s_1,t_1)$ and $H_2 \subset \SO(s_2,t_2)$ are the
holonomy groups of $M_1$ and $M_2$ respectively.  In particular this
means that the holonomy representation, by which we mean the
representation of the holonomy group on the tangent vectors, is
reducible.

In the riemannian case ($t=0$) there is a converse to this result,
known as the de~Rham decomposition theorem \cite{dR}.  This theorem
states that if the holonomy representation of a simply-connected
riemannian manifold is reducible, then the manifold is isometric to a
product of riemannian manifolds.  This allows one to restrict oneself
to groups acting irreducibly, and leads after some hard work to the
famous classification of riemannian holonomy groups of Cartan, Berger,
Simons and others.

The first sign that things do not quite work the same way in the
lorentzian ($t=1$) case is the fact that no proper subgroup of the
Lorentz group acts irreducibly on Minkowski spacetime.  Luckily this
does not mean that there are no interesting lorentzian holonomy
groups, only that asking for irreducibility is too strong.  In fact,
the lorentzian analogue of the decomposition theorem, due to Wu
\cite{Wu}, requires a weaker notion than irreducibility.

Recall that a representation of a group $H$ on a vector space $T$ is
reducible if there is a proper subspace $U \subset T$ which is
preserved by $H$.  Suppose in addition that $H$ preserves a
non-degenerate inner product on $T$; that is, $H$ acts on $T$ via a
(pseudo)unitary representation.  We say that the sub-representation
$U\subset T$ is \emph{non-degenerate} if the inner product is
non-degenerate when restricted to $U$.  In such a situation $T = U
\oplus U^\perp$, where $U^\perp$ is also a sub-representation.  If
this the case, then we say that $T$ is \emph{decomposable}.  Because
$U \cap U^\perp$ consists of vectors with zero norm, this can only be
nonzero for an indefinite inner product.  In other words, a unitary
representation is reducible if and only if it is decomposable.  If the
inner product is indefinite, so that the representation is
pseudo-unitary, then there is a distinction and one can have
representations which are reducible yet indecomposable.  It is the
stronger requirement of decomposability which is crucial for Wu's
version of the decomposition theorem.  Wu's theorem \cite{Wu} states
that if the holonomy representation is decomposable, then the manifold
is isometric to a product.  This fact makes the classification of
lorentzian holonomy groups a much harder problem than that of the
riemannian holonomy groups, since we cannot restrict ourselves to
groups acting irreducibly, but must also consider those groups acting
reducibly but indecomposably.  In fact, the problem has not been
completely solved in the classical case of four dimensions
\cite{Besse}.

It is precisely this exotic class of holonomy groups, acting reducibly
but indecomposably, which will be relevant in our construction of
supersymmetric $\M$-waves, to which we devote the rest of the paper.

\section{Supersymmetric gravitational $pp$-waves}
\label{sec:pp}

In this section we discuss gravitational $pp$-waves, since they are
the main ingredient in the construction to be presented below of
metrics with holonomy given in Table~\ref{tab:mwaves}.  We will
discuss the metrics of Brinkmann and of Bryant, and how to lift
gravitational $pp$-waves to supersymmetric $\M$-waves.

\subsection{Brinkmann metrics with parallel spinors}

By definition a gravitational $pp$-wave is a spacetime admitting a
parallel null vector.  As we discussed in the previous section,
contrary to the riemannian case, the existence of such a vector does
not necessarily split the spacetime.  The most general $d$-dimensional
lorentzian metric admitting a parallel null vector was written down by
Brinkmann \cite{Brinkmann2} almost 75 years ago.  In a coordinate
chart $(x^+,x^-,x^i)$, where $i=1,\dots,d{-}2$, adapted to the null
vector $\d_-$, the metric is given by
\begin{equation}
  \label{eq:brinkmann}
  ds^2 = 2 dx^+ dx^- + a\, (dx^+)^2 + b_i dx^i dx^+ + g_{ij} dx^i
  dx^j~,
\end{equation}
where $\d_- a = \d_- b_i = \d_- g_{ij} = 0$.  The holonomy group of
such a metric is contained in
\begin{equation*}
 \ISO(d-2) = \SO(d-2) \ltimes \RR^{d-2} \subset \SO(d-1,1)~,
\end{equation*}
which is the isotropy group of a null vector.

We are actually interested in those metrics whose holonomy group
reduces to the subgroup of (the spin cover of) $\ISO(d-2)$ which
preserves a spinor.  We will call such a spacetime a
\emph{supersymmetric Brinkmann wave}.  Since we are interested in
non-static spacetimes, we will consider only those subgroups which do
not leave any time-like vector invariant.  Maximal such subgroups are
of the form
\begin{equation}
  \label{eq:spinisotropy}
  K \ltimes \RR^{d-2}~,
\end{equation}
where $K\subset \Spin(d-2)$ preserves a spinor.  (If we were to
consider those subgroups which also leave a time-like vector
invariant, we would have only $K\subset \Spin(d-2)$.  Spacetimes with
holonomy contained in $K$ are static and hence decomposable into a
metric product.) A list of possible maximal subgroups $K \ltimes
\RR^{d-2}$ is given in Table~\ref{tab:spinisotropy}.  It is easy to
obtain other subgroups, not just the maximal ones.  Every group $H =
K \ltimes \RR^{d-2}$ appearing in dimension $d$ can be ``lifted'' to a
group $H \times \RR^{i}$ in dimension $d+s$, where $i=0,1,\dots,s$.
If we insist in the holonomy group acting indecomposably then we have
to choose $i=s$.  This procedure gives rise to the entries in the
table corresponding to $d=7$ and $d=11$, for example.

\begin{table}[h!]
\centering
\renewcommand{\arraystretch}{1.3}
\begin{tabular}{|>{$}r<{$}|>{$}c<{$}|}\hline
  d & H \subset \Spin(d-1,1) \\
  \hline\hline
  \leq 5 & \RR^{d-2} \\[3pt]
  6 & \Sp(1) \ltimes \RR^4 \\[3pt]
  7 & \left(\Sp(1) \ltimes \RR^4\right) \times \RR \\[3pt]
  8 & \SU(3) \ltimes \RR^6 \\[3pt]
  9 & G_2 \ltimes \RR^7 \\[3pt]
  10 & \Spin(7) \ltimes \RR^8\\[3pt]
  11 & \left( \Spin(7) \ltimes \RR^8\right) \times \RR \\
  \hline
\end{tabular}
\vspace{8pt}
\caption{Possible holonomy groups of $d$-dimensional supersymmetric
Brinkmann waves.  Only maximal subgroups are shown.}
\label{tab:spinisotropy}
\end{table}

The construction described in Section~\ref{sec:lift} starts with a
$d$-dimensional supersymmetric Brinkmann wave and warps it with a
riemannian manifold of dimension $11-d$.  It is therefore important to
be able to write down explicit metrics with holonomy groups in the
table.  We will write the most general local metric in dimension
$d\leq 5$ with holonomy given by $\RR^{d-2}$.  For $d\geq 6$ it is
possible to write down the most general local metric, but the
expression is not completely explicit.  In the case of $d=11$ the most
general such metric has been written down recently by Bryant
\cite{Bryant-spinors}.  His construction also works for $d=10$, and
indeed, after some modification, for $d=6,7,8,9$.  Bryant's
construction is reviewed briefly below.

\subsection{Supersymmetric Brinkmann waves in $d\leq 5$}
\label{sec:susywaves}

The strategy to construct the most general supersymmetric Brinkmann
wave is clear: we start with the most general $d$-dimensional
Brinkmann metric \eqref{eq:brinkmann} and impose that the holonomy be
contained in the appropriate subgroup $H\subset \Spin(d-1,1)$ in
Table~\ref{tab:spinisotropy}.  For $d\leq 5$ this means $H =
\RR^{d-2}$, and in this case the metric can be solved explicitly in
terms of a number of unknown functions of $x^+$.  In $d\geq 6$ this is 
not possible; for example, to write down the most general metric in
$d=6$, one would have to know the most general four-dimensional
hyperkähler metric.

A necessary condition for the spacetime to have holonomy $\RR^{d-2}$ is
that the metric be Ricci-null, so that condition \eqref{eq:riccinull}
is satisfied.  This is not a sufficient condition, but it is often an
easy condition to write down when trying to construct such a metric.

We will first of all show that the Ricci-null condition forces all
components of the Ricci tensor to vanish except for $R_{++}$.  Indeed,
because $\d_-$ is parallel, we have that $R_{a-} = 0$ for all $a$,
whence \emph{a priori} the only nonzero components of the Ricci tensor
are $R_{ij}$, $R_{++}$ and $R_{i+}$, or equivalently, $R_+{}^-$,
$R_+{}^i$, $R_i{}^-$ and $R_i{}^j$.  Now we use the Ricci-null
condition: $R_{ab} R_{ac} g^{bc} = 0$ for all $a$.  For $a=+$ this
relation tells us that $R_{+i} R_{+j} g^{ij} = 0$, whence $R_{+i} =
0$.  For $a=i$, we find that $R_{ij} R_{ik} g^{jk} = 0$, whence
$R_{ij} = 0$.  For $a=-$ the condition is vacuously satisfied.  In
summary, only $R_{++}$ can be nonzero.

\sssection{$d=2$}

The two-dimensional Brinkmann metric has the form
\begin{equation*}
  ds^2 = 2 dx^+ dx^- + a(x^+) (dx^+)^2~.
\end{equation*}
The holonomy is trivial, so that the metric is flat.  In fact, if we
change variables to $\Tilde x^+ = x^+$ and $\Tilde x^-= x^- + \half
f(x^+)$, where $f'= a$, then the metric is simply (dropping tildes)
\begin{equation}
  \label{eq:susybrinkmann2d}
  ds^2 = 2 dx^+ dx^-~.
\end{equation}

\sssection{$d=3$}

The three-dimensional Brinkmann metric has the form
\begin{equation*}
  ds^2 = 2 dx^+ dx^- + a (dx^+)^2 + b dx^+ dx^1 + c^2 (dx^1)^2~,
\end{equation*}
where $a,b,c$ are functions of $x^+$ and $x^1$.  We can simplify the
metric by using local diffeomorphisms.  We can change variables $x^1
\mapsto \Tilde x^1(x^+, x^1)$ such that $\d_1 \Tilde x^1 = c$.
Similarly we can define $\Tilde x^- = x^- + \half \phi(x^+,x^1)$,
where $\d_1 \phi = b - 2c \d_+ \Tilde x^1$.  Finally let $\Tilde x^+ =
x^+$.  In terms of the tilded variables (but dropping the tildes) the
metric becomes
\begin{equation}
  \label{eq:susybrinkmann3d}
  ds^2 = 2 dx^+ dx^- + a (dx^+)^2 + (dx^1)^2~,
\end{equation}
where $a$ is such that $\d_- a = 0$, but is otherwise arbitrary.

The holonomy is correct because in three dimensions the isotropy of a
null vector coincides with the isotropy of a null spinor: $\ISO(1)
\cong \RR \subset \Spin(2,1)$.

The Ricci tensor can be computed in terms of the function $a$.  From
the general discussion above we know that only $R_{++}$ will be
nonzero.  A simple calculation shows that $R_{++}=-\half \d_1^2 a$.
As we will see in Section~\ref{sec:ricciflat}, we will be able to
build $\M$-theory vacua out of this Brinkmann wave provided that
$R_{++}$ is only a function of $x^+$.  This means that $a$ is at most
quadratic in $x^1$.  Such waves are known as \emph{exact plane
waves}.

\sssection{$d=4$}

The four-dimensional Brinkmann metric has the form
\begin{equation}
  \label{eq:brinkmann4d}
  ds^2 = 2 dx^+ dx^- + a (dx^+)^2 + b_i dx^i dx^+ + g_{ij} dx^i dx^j~,
\end{equation}
where $i,j$ run from $1$ to $2$, and $a$, $b_i$ and $g_{ij}$ are
independent of $x^-$.  We can think of $g_{ij}$ as an $x^+$-dependent
family of two-dimensional riemannian metrics.  Because in two
dimensions every metric is conformally flat, we can assume that
$g_{ij} = c^2 \delta_{ij}$, where $c$ is independent of $x^-$.  The
metric is then
\begin{equation*}
  ds^2 = 2 dx^+ dx^- + a (dx^+)^2 + b_i dx^i dx^+ + c^2 dx^i dx^i~.
\end{equation*}
For generic $a$, $b_i$ and $c$ the holonomy of this metric is
contained in $\ISO(2) = \SO(2) \ltimes \RR^2 \subset \SO(3,1)$.  We
would like to restrict these functions so that the holonomy is
precisely $\RR^2$.

To simplify the calculations, we first impose the condition that the
spacetime be Ricci-null, so that only $R_{++}$ is different from
zero.  In other words, we set $R_{i+}$ and $R_{ij}$ equal to zero.

We will compute the Christoffel symbols $\Gamma_{ab}{}^c$ defined by
\begin{equation}
  \label{eq:christoffel}
  \nabla_a \d_b = \Gamma_{ab}{}^c\, \d_c~,
\end{equation}
where $\nabla$ is the Levi-Cività connection.  They can be read off
from
\begin{align*}
  \nabla_+ \d_+ &= \half \d_+ a \d_- - \half c^{-2} \d_i a \d_i +
  c^{-2} \d_+ b_i \d_i\\
  \nabla_+ \d_i &= \half \d_i a \d_- + \d_+\log c \d_i + \half c^{-2}
  (\d_i b_j - \d_j b_i) \d_j\\
  \nabla_i \d_j &= \Gamma_{ij}{}^k \d_k + \half (\d_i b_j + \d_j b_i)
  \d_- - c \d_+ c \delta_{ij} \d_-~,
\end{align*}
where $\Gamma_{ij}{}^k$ are the Christoffel symbols for the metric
$c^2 \delta_{ij}$.

In our conventions, the Riemann curvature tensor is defined by
\begin{equation}
  \label{eq:riemann}
  R_{abc}{}^d\, \d_d = -\left[ \nabla_a, \nabla_b\right]\, \d_c~,
\end{equation}
and the Ricci curvature is given by the following contraction:
\begin{equation}
  \label{eq:riccitensor}
  R_{ab} = R_{acb}{}^c~.
\end{equation}

Therefore the contribution to $R_{ij}$ comes from
\begin{equation*}
  R_{ij} = R_{iaj}{}^a = R_{i+j}{}^+ + R_{i-j}{}^- + R_{ikj}{}^k~.
\end{equation*}
Because $\d_-$ is parallel, $R_{-ab}{}^c = R_{abc}{}^+ = 0$, so that
$R_{ij} = R_{ikj}{}^k$; that is, $R_{ij}$ is the Ricci tensor of the
transverse metric $c^2 \delta_{ij}$.  The vanishing of $R_{ij}$ then
says that the transverse metric is Ricci-flat.  In two dimensions, a
Ricci-flat metric is flat, so that without loss of generality we can
take $c$ to be a function only of $x^+$.

We can then perform a diffeomorphism: $\Tilde x^i = c(x^+) x^i$ to
absorb the conformal factor $c^2$.  In the new variables, the metric
becomes
\begin{equation*}
  ds^2 = 2 dx^+ dx^- + a (dx^+)^2 + b_i dx^i dx^+ + dx^i dx^i~,
\end{equation*}
where $a$ and $b_i$ are arbitrary functions of $x^i$ and $x^+$.

The Levi-Cività connection for this metric can be read from the one
for the metric \eqref{eq:brinkmann4d}, by setting $c=1$ and
$\Gamma_{ij}{}^k = 0$:
\begin{align*}
  \nabla_+ \d_+ &= \half \d_+ a \d_- - \half \d_i a \d_i + \d_+ b_i
  \d_i\\
  \nabla_+ \d_i &= \half \d_i a \d_- + \half (\d_i b_j - \d_j b_i)
  \d_j\\
  \nabla_i \d_j &= \half (\d_i b_j + \d_j b_i) \d_-~.
\end{align*}

Next we set $R_{i+}=0$.  Only $R_{ij+}{}^k$ contributes to this
component of the Ricci tensor.  We compute
\begin{equation*}
 -[\nabla_i,\nabla_j] \d_+ = \half \d_k (\d_i b_j - \d_j b_i) \d_k
 \pmod {\d_-} ~,
\end{equation*}
whence
\begin{equation*}
  R_{ij+}^k = \half \d_k (\d_i b_j - \d_j b_i) = \half \epsilon_{ij}
  \d_k h~,
\end{equation*}
where $h = \d_1 b_2 - \d_2 b_1$.  The vanishing of $R_{i+}$ is then
simply
\begin{equation*}
 \d_k h = 0~,
\end{equation*}
so that $\d_1 b_2 - \d_2 b_1$ is only a function of $x^+$.  The most
general solution to this equation is
\begin{equation*}
  b_i = \epsilon_{ij} x^j b(x^+) + \d_i \phi~,
\end{equation*}
for an arbitrary function $\phi$ independent of $x^-$.

We can reabsorb the gradient term in $b_i$ by a local diffeomorphism
$\Tilde x^- = x^- + \half \phi$.  In terms of the new variables
(dropping tildes) the most general four-dimensional Ricci-null
Brinkmann metric becomes
\begin{equation}
  \label{eq:susybrinkmann4d}
  ds^2 = 2 dx^+ dx^- + a (dx^+)^2 + b \epsilon_{ij} x^j dx^i dx^+ +
  dx^i dx^i~,
\end{equation}
where $b=b(x^+)$ and $a=a(x^+,x^i)$ are arbitrary functions.

We claim that this metric already has the correct holonomy.  Let us
introduce the following pseudo-orthonormal coframe:
\begin{equation*}
  \btheta^{\Hat +} = dx^+~,\quad
  \btheta^{\Hat -} = dx^- + \half a dx^+ + \half b \epsilon_{ij}
  x^j dx^i~,
  \quad\text{and}\quad
  \btheta^{\Hat i} = dx^i~,
\end{equation*}
relative to which the metric becomes
\begin{equation*}
  ds^2 = 2 \btheta^{\Hat +} \btheta^{\Hat -} + \btheta^{\Hat
  i} \btheta^{\Hat i}~.
\end{equation*}

The connection one-form can be computed from the first structure
equation:
\begin{equation}
  \label{eq:fse}
  d\btheta^{\Hat a}  = \btheta^{\Hat b} \wedge
  \bomega_{\Hat b}{}^{\Hat a}~.
\end{equation}
One finds the following nonzero components:
\begin{align*}
  \bomega_{\Hat i}{}^{\Hat j} &= b \epsilon_{ij} dx^+~,\\
  \text{and}\quad \bomega_{\Hat +}{}^{\Hat j} &= -\half \d_j a
  dx^+ + b' \epsilon_{ji} x^i dx^+ + b \epsilon_{ij} dx^i~,
\end{align*}
where $b'=\d_+ b$.

The curvature two-form can be computed using the second structure
equation:
\begin{equation}
  \label{eq:sse}
  \bOmega_{\Hat a}{}^{\Hat b} =
  d \bomega_{\Hat a}{}^{\Hat b} -
  \bomega_{\Hat a}{}^{\Hat c} \wedge
  \bomega_{\Hat c}{}^{\Hat b}~.
\end{equation}
One finds that the $\so(2)$ component of the curvature vanishes:
\begin{align*}
  \bOmega_{\Hat i}{}^{\Hat j} = 0~.
\end{align*}
Using the Ambrose--Singer theorem this says that the $\so(2)$ part of
the holonomy algebra $\fh$ vanishes, leaving $\fh \cong \RR^2$.

Of course, the forms for the metrics are not unique.  By a change of
coordinate it is possible to eliminate the mixed $dx^i dx^+$ terms at
the cost of introducing some conformal factor in the transverse metric 
(see, for example, \cite{Arkady}).

Finally we record the Ricci tensor of the metric
\eqref{eq:susybrinkmann4d}, whose only nonzero component is
\begin{equation*}
  R_{++} = -\half \bigtriangleup a + 2 b^2~,
\end{equation*}
where $\bigtriangleup = \d_i\d_i$ is the laplacian on functions of the
transverse  coordinates $x^i$.  This will be a function of $x^+$ alone 
provided that $\bigtriangleup a$ has no dependence on $x^i$.  Clearly
there are plenty of such functions.  We will see in
Section~\ref{sec:lift} how any of these metrics can be used as an
ingredient in a supersymmetric $\M$-wave.

\sssection{$d=5$}

The five-dimensional Brinkmann metric has the form
\begin{equation}
  \label{eq:brinkmann5d}
  ds^2 = 2 dx^+ dx^- + a (dx^+)^2 + b_i dx^i dx^+ + g_{ij} dx^i dx^j~,
\end{equation}
where now $i,j$ run from $1$ to $3$, and $a$, $b_i$ and $g_{ij}$ are
independent of $x^-$.  We can again think of $g_{ij}$ as an
$x^+$-dependent family of three-dimensional riemannian metrics.  For
generic choices of $a$, $b_i$ and $g_{ij}$, the holonomy of this
metric is contained in $\ISO(3) = \SO(3) \ltimes \RR^3 \subset
\SO(4,1)$.  We would like to choose $a$, $b_i$ and $g_{ij}$ in such a
way that the holonomy is simply $\RR^3$.  This way the spacetime will
admit a parallel null spinor.  In particular it will be Ricci-null, so
that $R_{ij} = R_{+i} = 0$.  As in the case of $d=4$, we start by
imposing the vanishing of $R_{ij}$, as imposing this condition from
the start will simplify subsequent calculations.

The Levi-Cività connection of the above metric is given by
\begin{align*}
  \nabla_+ \d_+ &= \half \d_+ a \d_- + g^{ij} (\d_+ b_j - \half \d_j
  a) \d_i\\
  \nabla_+ \d_i &= \half \d_i a \d_- + \half g^{jk} \left( \d_+ g_{ik} 
    + \d_i b_k - \d_k b_i \right) \d_j\\
  \nabla_i \d_j &= \Gamma_{ij}{}^k \d_k + \half \left( \d_i b_j + \d_j 
  b_i - \d_+ g_{ij} \right) \d_-~,
\end{align*}
where $\Gamma_{ij}{}^k$ are the Christoffel symbols for the transverse 
metric $g_{ij}$.

The only contribution to $R_{ij}$ again comes from the curvature
tensor of the transverse space; whence setting $R_{ij}=0$ means that
the transverse metric $g_{ij}$ must be Ricci-flat.  In three
dimensions, a riemannian metric is Ricci flat if and only if it is
flat.  This means that we can choose local coordinates so that
$g_{ij} = c^2(x^+) \delta_{ij}$.

As before we can perform a local diffeomorphism: $\Tilde x^i = c(x^+)
x^i$ to absorb the conformal factor $c^2$.  In the new variables, the
metric becomes
\begin{equation*}
  ds^2 = 2 dx^+ dx^- + a (dx^+)^2 + b_i dx^i dx^+ + dx^i dx^i~,
\end{equation*}
where $a$ and $b_i$ are arbitrary functions of $x^i$ and $x^+$.

Unlike the four-dimensional case, the vanishing of $R_{+i}$ is not
sufficient to restrict the holonomy.  To restrict the holonomy we will 
compute the curvature two-form and impose the vanishing of the
$\so(3)$ components directly.

To this effect we introduce the pseudo-orthonormal coframe:
\begin{equation*}
  \btheta^{\Hat +} = dx^+~,\quad
  \btheta^{\Hat -} = dx^- + \half b_i dx^i + \half a dx^+~,
  \quad\text{and}\quad
  \btheta^{\Hat i} = dx^i~.
\end{equation*}
The connection one-form can be computed from the first structure
equation \eqref{eq:fse}.  One finds the following nonzero components:
\begin{align*}
  \bomega_{\Hat i}{}^{\Hat j} &= \tfrac14 (\d_i b_j - \d_j b_i) 
  dx^+~,\\
  \text{and}\quad
  \bomega_{\Hat +}{}^{\Hat j} &= \half (\d_+ b_j - \d_j a) dx^+
  - \tfrac14 (\d_j b_i - \d_i b_j) dx^i~.
\end{align*}

The curvature two-form can be computed using the second structure
equation \eqref{eq:sse}.  One finds that the $\so(3)$ component of the
curvature is given by
\begin{align*}
  \bOmega_{\Hat i}{}^{\Hat j} = -\tfrac14 \d_k \left( \d_i b_j
  - \d_j b_i \right) dx^k \wedge dx^+~.
\end{align*}
By the Ambrose--Singer theorem this is the $\so(3)$ components of the
holonomy algebra, which must vanish.  As in the four-dimensional case
this is equivalent to
\begin{equation*}
  \d_k \left( \d_i b_j - \d_j b_i \right) = 0~,
\end{equation*}
whose most general solution is
\begin{equation*}
  b_i = \epsilon_{ijk} x^j f_k(x^+) + \d_i \phi~,
\end{equation*}
where the gradient term can again be reabsorbed via a local
diffeomorphism $\Tilde x^- = x^- + \half \phi$.

In summary, the most general supersymmetric Brinkmann wave in five
dimensions has the metric
\begin{equation}
  \label{eq:susybrinkmann5d}
  ds^2 = 2 dx^+ dx^- + a (dx^+)^2 + \epsilon_{ijk} x^j f_k dx^i dx^+ +
  dx^i dx^i~,
\end{equation}
where $f_j=f_j(x^+)$ and $a=a(x^+,x^i)$ are arbitrary functions.

Finally we record the Ricci tensor corresponding to this metric
\eqref{eq:susybrinkmann5d}, whose only nonzero component is
\begin{equation*}
  R_{++} = -\half \bigtriangleup a - 2 f^2~,
\end{equation*}
where $\bigtriangleup$ is the laplacian on functions of the transverse
coordinates $x^i$ and $f^2 = f_i f_i$.  As in the $d=3,4$ cases
treated above, we see that the condition that $R_{++}$ depend only on
$x^+$ can be met by a large class of functions $a$, and for each such
function we will see in the Section~\ref{sec:lift}, how to construct a
supersymmetric $\M$-wave by an appropriate warping construction.

\subsection{The Bryant metrics for $d=11$ and $d=10$}

Bryant \cite{Bryant-spinors} has written down the most general local
metric admitting parallel null spinors in $d=11$, and his results are
easily adapted to $d=10$, and indeed to lower dimensions.  The most
general metric in $d=11$ admitting a parallel null spinor is given by
specialising the Brinkmann metric \eqref{eq:brinkmann}.  The Bryant
metric is given by
\begin{equation}
  \label{eq:bryant}
  ds^2 = 2 dx^+ dx^- + a\, (dx^+)^2 + (dx^9)^2 + g_{ij} dx^i dx^j~,
\end{equation}
where $i,j$ now run from $1$ to $8$, $\d_- a = 0$ and now $g_{ij}$ is
an $x^+$-dependent family of metrics with holonomy contained in
$\Spin(7)$ and obeying the property that
\begin{equation}
  \label{eq:casd}
  \d_+ \Omega = \lambda \Omega + \Psi~,
\end{equation}
where $\Omega$ is the self-dual $\Spin(7)$-invariant Cayley $4$-form,
$\lambda$ a smooth function of $(x^+,x^i)$ and $\Psi$ an anti-self
dual $4$-form.  Following Bryant we call such a family of metrics
\emph{conformal anti-self dual}.  This condition is locally trivial,
in the sense that one can use diffeomorphisms to make sure that any
one-parameter family of $\Spin(7)$ holonomy metrics is conformal
anti-self dual \cite{Bryant-spinors}.

Special cases of this metric have already appeared in the literature
in the context of solutions to supergravity or superstring theory.
Already in \cite{Mwave} there is mention of unpublished work of
J~Richer concerning the metric
\begin{equation}
  \label{eq:richer}
  ds^2 = 2dx^+ dx^- + a\, (dx^+)^2 + \sum_{i=1}^9 dx^i dx^i~,
\end{equation}
where $a$ obeys $\d_- a = 0$ but is otherwise arbitrary.  This is the
special case of the Bryant metric \eqref{eq:bryant} where $g_{ij} =
\delta_{ij}$.  For generic $a$, this metric has holonomy $\RR^9$.

We can gain some insight into the conformal anti-self dual condition
\eqref{eq:casd} by constructing some examples.  One such class of
examples is the one in which the $x^+$ dependence of the metric is via
a conformal factor, which means $\Psi =0$ in the above expression.  In
other words, take
\begin{equation}
  \label{eq:conformal}
  g_{ij}(x^+,x^i) = e^{2\sigma(x^+)} \, \bar g_{ij}(x^i)~,
\end{equation}
where $\bar g_{ij}$ is a fixed metric whose holonomy is contained in
$\Spin(7)$.  (More generally we could also let $\sigma$ depend on
$x^i$, but then $\bar g_{ij}$ is only conformal to a metric with
holonomy in $\Spin(7)$.)  In Section~\ref{sec:warp3d} we will
construct many classes of Bryant metrics, realising the holonomy
groups in Table~\ref{tab:mwaves} and satisfying the the conformal
anti-self dual condition in a more non-trivial way than the one just
described.

The Bryant metric \eqref{eq:bryant} will be a supersymmetric vacuum of
$d{=}11$ supergravity if and only if it is Ricci-flat.  As explained
above, this is not automatic since the existence of a parallel null
spinor only implies the weaker condition \eqref{eq:riccinull} that the
spacetime be Ricci-null.  As we saw above, this forces all components
of the Ricci tensor to vanish except for $R_{++}$.  Therefore the
vanishing of $R_{++}$ becomes a differential equation for the unknown
function $a$.

Let us write this equation down explicitly in the special case of the
Bryant metric \eqref{eq:bryant} satisfying equation
\eqref{eq:conformal}.

We will therefore consider the following eleven-dimensional metric:
\begin{equation}
  \label{eq:bryantconformal}
  ds^2 = 2 dx^+ dx^- + a\, (dx^+)^2 + (dx^9)^2 + e^{2\sigma(x^+)}
  g_{ij} dx^i dx^j~,
\end{equation}
where $\d_- a = 0$ and $g_{ij}$ is independent of $(x^\pm, x^9)$.  The
nonzero Christoffel symbols can be read from equation
\eqref{eq:christoffel} and the following relations:
\begin{gather*}
  \nabla_+ \d_+ = \half \d_+ a\, \d_- - \half \d_9 a\, \d_9 - \half
  e^{-2\sigma} g^{ij} \d_i a\, \d_j\\
  \nabla_+ \d_9 = \nabla_9 \d_+ = \half \d_9 a\, \d_-\\
  \nabla_+ \d_i = \nabla_i \d_+ = \half \d_i a\, \d_- + \sigma'n\,
  \d_i\\
  \nabla_i \d_j = \Gamma_{ij}{}^k\, \d_k - e^{2\sigma} g_{ij}
  \sigma'\, \d_-~,
\end{gather*}
where $\Gamma_{ij}{}^k$ are the Christoffel symbols of the metric
$g_{ij}$, and $\sigma' = \d_+\sigma$.

Using the above Christoffel symbols one computes the Riemann
curvature tensor \eqref{eq:riemann} and finds the following nonzero
components:
\begin{gather*}
  R_{+9+}{}^9 =   R_{9+9}{}^- = -\half \d_9^2 a\\
  R_{+9i}{}^- = R_{+i9}{}^- = R_{i++}{}^9 = \half \d_9\d_i a\\
  R_{+9+}{}^i = -\half e^{-2\sigma} g^{ij}\, \d_9\d_j a\\
  R_{+i+}{}^j = - \half \nabla_i \d^j a - \left(\sigma'' +
  (\sigma')^2\right) \delta_i{}^j \\
  R_{+ij}{}^- = \half \nabla_i\d_j a + \left(\sigma'' +
  (\sigma')^2\right) e^{2\sigma}\, g_{ij}~,
\end{gather*}
together with $R_{ijk}{}^\ell$.  We now compute the Ricci curvature,
defined by \eqref{eq:riccitensor}.  Using the fact that $\Spin(7)$
holonomy manifolds are Ricci-flat we get that $R_{ij}=0$.  Similarly,
one sees that all other components vanish except for
\begin{equation*}
  R_{++}= -\half \d_9^2 a - \half \nabla_i \d^i a - 8 \left( \sigma''
  + (\sigma')^2\right)~,
\end{equation*}
as expected form the fact that $ds^2$ admits a parallel null spinor.
Setting $R_{++}$ to zero yields a differential equation for $a$:
\begin{equation}
  \label{eq:ricciflat}
  \d_9^2 a + \nabla_i \d^i a = - 16 \left( \sigma'' +
  (\sigma')^2\right)~.
\end{equation}
Assuming the right-hand side is non-vanishing, we can simplify the
equation by writing the function $a$ in the form
\begin{equation*}
a(x^+, x^9, x^i) = -16 \left(\sigma'' + (\sigma')^2 \right) b(x^9,
x^i)~,
\end{equation*}
we see that equation \eqref{eq:ricciflat} becomes
\begin{equation*}
\bigtriangleup b = 1~,
\end{equation*}
where $\bigtriangleup$ is the laplacian on functions of $(x^9,x^i)$.

Dropping all mention of $x^9$, which means taking $a$ to be
independent of $x^9$ and performing a dimensional reduction on this
coordinate, we arrive at the most general ten-dimensional metric with
holonomy $\Spin(7) \ltimes \RR^8$:
\begin{equation}
  \label{eq:bryant10d}
  ds^2 = 2 dx^+ dx^- + a\, (dx^+)^2 + g_{ij} dx^i dx^j~,  
\end{equation}
where $a=a(x^+,x^i)$ is an arbitrary function, and $g_{ij}$ is a
conformal anti-self dual family of metrics whose holonomy is contained 
in $\Spin(7)$.

Special cases of this metric have been studied in the literature, in
the context of wave-like solutions to type IIA supergravity and
superstring theories \cite{BKO,HoTs,GSR}.

Similarly one can write local metrics for supersymmetric Brinkmann
waves in lower dimension, provided that we choose the family of
metrics $g_{ij}(x^+)$ (with $i,j = 1, \cdots, d-2$) to have the
appropriate holonomy and that the variation under $x^+$ is suitably
constrained.  (Compare the appendix in \cite{HoTs} for some examples
of such spacetimes.)

\section{Lifting the Brinkmann waves}
\label{sec:lift}

In this section we describe a construction of supersymmetric
$\M$-waves obtained by lifting $d$-dimensional supersymmetric
Brinkmann waves to eleven dimensions.  The idea is to warp the
Brinkmann wave with an ($11-d$)-dimensional riemannian manifold
admitting parallel spinors.  The resulting spacetime is indecomposable
and admits a parallel null spinor.  If the Brinkmann wave satisfies
the additional requirement that $R_{++}$ only depends on $x^+$, then
it will be possible to choose the warp factor so that the warped
product is Ricci-flat, and hence a supersymmetric $\M$-wave.

\subsection{Warped products and Brinkmann waves}

Let $B_d$ be a $d$-dimensional supersymmetric Brinkmann wave.  Its
holonomy group $H(B)\subset \Spin(d-1,1)$ is listed in
Table~\ref{tab:spinisotropy} or is built from them according to the
procedure outlined above that table.  Let $K$ be an
($11-d$)-dimensional riemannian manifold admitting parallel spinors.
We will take both $B$ and $K$ to be simply-connected for simplicity,
but clearly this does not represent any real loss of generality, at
least conceptually; although in the non-simply connected case some
further supersymmetry might be broken.

Let $(x^a) = (x^+,x^-,x^i)$, for $i=1,2,\dots,d-2$, be local
coordinates for the Brinkmann wave $B$, and let $(y^m)$ for
$m=1,2,\dots,11-d$ be local coordinates for $K$.  Let $\sigma$ be an
arbitrary function of $x^+$.  The warped product $M = B \times_\sigma
K$ is topologically $B \times K$, but not metrically.  Instead the
metric on $M$ is given by
\begin{equation}
  \label{eq:Mwave}
  ds^2 = g_{ab}(x) dx^a dx^b + e^{2\sigma} \, h_{mn}(y) dy^m dy^n~,
\end{equation}
where $g_{ab}$ is the Brinkmann metric and $h_{mn}$ is the metric on
$K$.

Table~\ref{tab:warpings} lists possible warped products of this type,
according to the holonomy groups $H(B)$ and $H(K)$.  Only the maximal
holonomy groups are listed.  For example, for $d{=}2,3$ we can have a
number of subgroups of $\Spin(7)$ as the holonomy group of $K$.  Some,
like $\SU(4)$ and $\Sp(2)$ do not split $K$, but some like $\Sp(1)
\times \Sp(1)$ do.  It is not important that $K$ be irreducible for
$M$ to be indecomposable, but it is important that $B$ be
indecomposable, at least for this construction.  This constrains the
subgroups which can appear as $H(B)$.  (Compare with the discussion
atop Table~\ref{tab:spinisotropy}.)

This construction can be generalised in the following way. If the
manifold $K$ is reducible, so that it is metrically a product $K = K_1
\times \dots \times K_N$, then we can use a different warp factor
$\sigma_i$ in each irreducible component $K_i$.  We shall not dwell on 
this generalisation, but we will see it again briefly in
Section~\ref{sec:warp3d}, where we will construct metrics with the
holonomies in Table~\ref{tab:mwaves}.

\begin{table}[h!]
\centering
\renewcommand{\arraystretch}{1.3}
\begin{tabular}{|>{$}r<{$}|>{$}c<{$}|>{$}c<{$}|>{$}c<{$}|}\hline
  d & H(B) \subset \Spin(d-1,1) & H(K) \subset \Spin(11-d) & 11-d \\
  \hline\hline
  2 & \{1\} & \Spin(7) & 9\\
  3 & \RR & \Spin(7) & 8 \\
  4 & \RR^2 & G_2 & 7 \\
  5 & \RR^3 & \SU(3) & 6 \\
  6 & \Sp(1) \ltimes \RR^4 & \Sp(1) & 5 \\
  7 & \left(\Sp(1) \ltimes \RR^4\right) \times \RR & \Sp(1) & 4\\
  8 & \SU(3) \ltimes \RR^6 & \{1\} & 3\\
  9 & G_2 \ltimes \RR^7 & \{1\} & 2\\
  10 & \Spin(7) \ltimes \RR^8 & \{1\} & 1\\
  \hline
\end{tabular}
\vspace{8pt}
\caption{Possible supersymmetric eleven-dimensional warped
  products $B \times_\sigma K$. Only maximal subgroups are shown on
  either side.}
\label{tab:warpings}
\end{table}

We claim that the metric \eqref{eq:Mwave} admits a parallel null
spinor.  Moreover we will see that when $B$ is such that the only
nonzero component $R_{++}$ of its Ricci tensor depens only on $x^+$,
it will be possible to choose the warping function $\sigma$ in such a
way that the metric \eqref{eq:Mwave} is Ricci flat, and thus providing
a supersymmetric $\M$-theory vacuum.  To see this we will compute the
holonomy algebra of the warped product and also the Ricci tensor.  We
will see that the holonomy algebra always agrees with the Lie algebra
of the holonomy groups in Table~\ref{tab:mwaves}, and that all such
Lie algebras appear in this way.

\subsection{The holonomy group of a warped product}
\label{sec:warhol}

We start by computing the holonomy algebra of the warped product $M$.
By the Ambrose--Singer theorem the holonomy algebra is generated by
the components of the curvature two-form, so we compute this.

Consider a pseudo-orthonormal coframe $\btheta^{\Hat a}$ for $B$ and an
orthonormal coframe $\btheta^{\Hat m}$ for $K$.  We form a
pseudo-orthonormal coframe $(\Bar\btheta^{\Hat A}) = (\Bar\btheta^{\Hat
  a}, \Bar\btheta^{\Hat m})$ for $M$, where $\Bar\btheta^{\Hat a} =
\btheta^{\Hat a}$ and $\Bar\btheta^{\Hat m} = e^\sigma \btheta^{\Hat m}$.
In terms of this coframe the metric \eqref{eq:Mwave} becomes
\begin{equation*}
  ds^2 = \eta_{\Hat a \Hat b} \Bar\btheta^{\Hat a} \Bar\btheta^{\Hat b}
  + \delta_{\Hat m \Hat n} \Bar\btheta^{\Hat m} \Bar\btheta^{\Hat n}~.
\end{equation*}

The connection one-form follows from the first structure equation:
\begin{align*}
  d \Bar\btheta^{\Hat a} &=
  \Bar\btheta^{\Hat b} \wedge \Bar\bomega_{\Hat b}{}^{\Hat a} +
  \Bar\btheta^{\Hat m} \wedge \Bar\bomega_{\Hat m}{}^{\Hat a}\\
  d \Bar\btheta^{\Hat m} &=
  \Bar\btheta^{\Hat n} \wedge \Bar\bomega_{\Hat n}{}^{\Hat m} +
  \Bar\btheta^{\Hat b} \wedge \Bar\bomega_{\Hat b}{}^{\Hat m}~.
\end{align*}
Computing the left-hand sides of these equations and using the first
structure equations in $B$ and $K$ separately, we find
\begin{equation*}
  \Bar\bomega_{\Hat b}{}^{\Hat a} = \bomega_{\Hat b}{}^{\Hat a}~,\quad
  \Bar\bomega_{\Hat m}{}^{\Hat n} = \bomega_{\Hat m}{}^{\Hat n}
  \quad\text{and}\quad
  \Bar\bomega_{\Hat a}{}^{\Hat m} =\d_{\Hat a} \sigma
  \Bar\btheta^{\Hat m}~,
\end{equation*}
where $\d_{\Hat a} \equiv e_{\Hat a}^a\d_a$ are the frames dual to the
coframe $\btheta^{\Hat a}$.

We compute the curvature two-form using the second structure equation:
\begin{align*}
  \Bar\bOmega_{\Hat a}{}^{\Hat b} &= d\Bar\bomega_{\Hat a}{}^{\Hat b} - 
  \Bar\bomega_{\Hat a}{}^{\Hat c} \wedge \Bar\bomega_{\Hat c}{}^{\Hat b} 
  - \Bar\bomega_{\Hat a}{}^{\Hat m} \wedge \Bar\bomega_{\Hat m}{}^{\Hat
    b}\\
  \Bar\bOmega_{\Hat a}{}^{\Hat m} &= d\Bar\bomega_{\Hat a}{}^{\Hat m} - 
  \Bar\bomega_{\Hat a}{}^{\Hat b} \wedge \Bar\bomega_{\Hat b}{}^{\Hat m} 
  - \Bar\bomega_{\Hat a}{}^{\Hat n} \wedge \Bar\bomega_{\Hat n}{}^{\Hat
    m}\\
  \Bar\bOmega_{\Hat m}{}^{\Hat n} &= d\Bar\bomega_{\Hat m}{}^{\Hat n} - 
  \Bar\bomega_{\Hat m}{}^{\Hat a} \wedge \Bar\bomega_{\Hat a}{}^{\Hat n} 
  - \Bar\bomega_{\Hat m}{}^{\Hat p} \wedge \Bar\bomega_{\Hat p}{}^{\Hat
    n}~.
\end{align*}
Using the explicit expressions of the connection one-forms, one finds
\begin{align*}
  \Bar\bOmega_{\Hat a}{}^{\Hat b} &= \bOmega_{\Hat a}{}^{\Hat b}\\
  \Bar\bOmega_{\Hat a}{}^{\Hat m} &= \left( \d_{\Hat a} \sigma \d_{\Hat
      b} \sigma \Bar\btheta^{\Hat b} + \d_{\Hat b} \d_{\Hat a} \sigma
    \Bar\btheta^{\Hat b} - \bomega_{\Hat a}{}^{\Hat b} \d_{\Hat b}
    \sigma\right) \wedge \Bar\btheta^{\Hat m}\\
  \Bar\bOmega_{\Hat m}{}^{\Hat n} &= \bOmega_{\Hat m}{}^{\Hat n} -
  \d_{\Hat a} \sigma \d^{\Hat a} \sigma \Bar\btheta^{\Hat m} \wedge
  \btheta^{\Hat n}~.
\end{align*}

Since $\sigma$ is only a function of $x^+$, its gradient $\d_{\Hat a}
\sigma$ is null: $\d_{\Hat a} \sigma \d^{\Hat a} \sigma = 0$.  This
sets $\Bar\bOmega_{\Hat m}{}^{\Hat n} = \bOmega_{\Hat m}{}^{\Hat n}$.
We now claim that the only nonzero component of $\Bar\bOmega_{\Hat
a}{}^{\Hat m}$ is $\Bar\bOmega_{\Hat +}{}^{\Hat m}$.  Postponing the
proof of this statement momentarily, let us see what this says about
the holonomy group of the warped product.

The Ambrose--Singer theorem tells us that the holonomy algebra at a
point $p\in M$ is the subalgebra of $\so(10,1)$ generated by the
elements $\Bar\bOmega^{\Hat A \Hat B}(p)$ under linear combinations and
Lie brackets.  These elements are the following: $\Bar\bOmega^{\Hat a
  \Hat b}$, which span the holonomy algebra $\fh_B$ of the Brinkmann
wave, $\Bar\bOmega^{\Hat m\Hat n}$ which span the holonomy algebra
$\fh_K$ of the transverse space $K$, and $\Bar\bOmega^{\Hat - \Hat m}$
which span an abelian algebra isomorphic to $\RR^{11-d}$.  The Lie
algebra $\fh_B$ is isomorphic to $\fk \ltimes \RR^{d-2}$, where $\fk
\subset \so(d-2)$ preserves some spinors and acts on $\RR^{d-2}$ by
restricting the vector representation of $\so(d-2)$ to $\fk$.  The
holonomy algebra $\fh_M$ is then isomorphic to $(\fk \times \fh_K)
\ltimes \RR^9$.

Going through the Lie algebras of the holonomy groups in
Table~\ref{tab:warpings} we see that the holonomy algebra of $M = B
\times_\sigma K$ is given by the Lie algebra of the following groups,
indexed according to the dimension $d$ of the Brinkmann wave:
\begin{itemize}
\item for $d=2,3,10,11$,
  \begin{equation*}
    \Spin(7) \ltimes \RR^9 \cong \left( \Spin(7) \ltimes \RR^8 \right) 
    \times \RR~;
  \end{equation*}
\item for $d=4,9$,
  \begin{equation*}
    G_2 \ltimes \RR^9 \cong \left( G_2 \ltimes \RR^7 \right) 
    \times \RR^2~;
  \end{equation*}
\item for $d=5,8$,
  \begin{equation*}
    \SU(3) \ltimes \RR^9 \cong \left( \SU(3) \ltimes \RR^6 \right) 
    \times \RR^3~;
  \end{equation*}
\item and for $d=6,7$,
  \begin{equation*}
    \left( \Sp(1) \times \Sp(1) \right) \ltimes \RR^9 \cong \left(
    \Sp(1) \ltimes \RR^4 \right) \times \left( \Sp(1) \ltimes \RR^4
    \right) \times \RR~.
  \end{equation*}
\end{itemize}

These holonomy groups appear in Table~\ref{tab:mwaves}.  The other
groups in that table can be obtained via the same construction, but
where we take spaces $B$ and $K$ which do not realise the maximal
allowed holonomy group, but a subgroup.  In Section~\ref{sec:warp3d}
below we will see how to obtain all the holonomy groups in
Table~\ref{tab:mwaves} by warping a three-dimensional Brinkmann wave
with holonomy $\RR$ together with an eight-dimensional space $K$ of
holonomy $H(K) \subseteq \Spin(7)$ to obtain a space with holonomy
$H(K) \ltimes \RR^9$.

Now we return to the proof of the statement that the only mixed
component of the curvature two-form is $\Bar\bOmega_{\Hat +}{}^{\Hat m}
= - \Bar\bOmega_{\Hat m}{}^{\Hat -}$.  This will follow from the
explicit expression for $\Bar\bOmega_{\Hat +}{}^{\Hat m}$ given above,
and the following two observations: that the only nonzero component of
the gradient $\d_{\Hat a} \sigma$ is $\d_{\Hat +} \sigma$, and that
$\bomega_{\Hat a}{}^{\Hat +} = 0$.

To prove these assertions we must go back to the metric
\eqref{eq:brinkmann} for a Brinkmann wave.  A convenient
pseudo-orthonormal coframe $(\btheta^{\Hat a})$ is given by
$(\btheta^{\Hat +}, \btheta^{\Hat -}, \btheta^{\Hat i})$, where
$\btheta^{\Hat i}$ is an orthonormal coframe for $g_{ij}$,
$\btheta^{\Hat +} = dx^+$ and
\begin{equation*}
  \btheta^{\Hat -} = dx^- + \half a dx^+ + \half b_i dx^i~.
\end{equation*}
From this expression it is plain to see that dual pseudo-orthonormal
frame $\d_{\Hat a} = e_{\Hat a}{}^a \d_a$ is such that the only
nonzero component of $e_{\Hat a}{}^+$ is $e_{\Hat +}{}^+ = 1$.  Since
the warp factor $\sigma$ is only a function of $x^+$, it follows that
$\d_{\Hat a} \sigma$ vanishes except for $\d_{\Hat +} \sigma = \d_+
\sigma$.  This proves the first assertion.

To prove the second assertion we consider the first structure equation
\eqref{eq:fse} relating the coframe with the connection one-forms.
Because $dx^-$ does not appear in $d\btheta^{\Hat a}$ for any $a$ and
$dx^-$ only appears in $\btheta^{\Hat -}$ it follows that $\btheta^{\Hat
  -}$ must not appear in the right-hand side of the structure equation
\eqref{eq:fse}.  In other words, $\bomega_{\Hat -}{}^{\Hat a} = 0$ for
all $a$.  But because the coframe is pseudo-orthonormal, one has that
$\bomega_{\Hat a}{}^{\Hat +} = - \eta_{\Hat a\Hat b} \bomega_{\Hat
  -}{}^{\Hat b} = 0$.

\subsection{Some metrics with holonomy in Table~\ref{tab:mwaves}}
\label{sec:warp3d}

In this section we will show how to construct indecomposable
eleven-dimensional spacetimes with each of the holonomy groups in
Table~\ref{tab:mwaves}.  The resulting spacetime will be a
(generalised) warped product $M$ of a three-dimensional supersymmetric
Brinkmann wave $B$ and an eight-di\-men\-sion\-al manifold $K$ with
holonomy contained in $\Spin(7)$.  This guarantees that $K$ and,
hence, $M$ admit a parallel spinor.  Equation
\eqref{eq:susybrinkmann3d} describes the most general local metric for
a three-dimensional supersymmetric Brinkmann wave.  It depends on one
arbitrary function $a$ of $x^+$ and $x^1$.  Since the
eight-di\-men\-sion\-al space $K$ need not be irreducible, we have
many possible choices.  We will now discuss each choice very briefly,
and for each choice we will write down a metric with holonomy group in
Table~\ref{tab:mwaves} depending on at least two functions: the
function $a$ in the metric for $B$ and one warping function for each
irreducible component of $K$.

\sssection{$\left(\Spin(7)\ltimes\RR^8\right)\times\RR$}

We take $K$ to be irreducible with holonomy $\Spin(7)$, and let the
metric be given by
\begin{equation}
  \label{eq:Kmetric}
  ds^2(K) = g_{mn}(y) dy^m dy^n~.
\end{equation}
Let $\sigma$ be an arbitrary function of $x^+$ and consider the metric
given by
\begin{equation}
  \label{eq:spin7}
  ds^2 = 2 dx^+ dx^- + a (dx^+)^2 + (dx^1)^2 + e^{2\sigma} g_{mn} dy^m 
  dy^n~.
\end{equation}
For generic $a$ and $\sigma$ the holonomy of this metric is the
maximal subgroup $G$ given in equation \eqref{eq:exotic}.  From the
appendix it follows that there is precisely one parallel spinor.  If
the manifold is Ricci-flat (see below) then it is an $\M$-wave
preserving precisely $\frac1{32}$ of the supersymmetry.

\sssection{$\left(\SU(4)\ltimes\RR^8\right)\times\RR$}

Let $K$ be irreducible with metric given by equation
\eqref{eq:Kmetric} but with holonomy $\SU(4)$.  Then the warped
product metric given by the expression \eqref{eq:spin7} now has
holonomy
\begin{equation*}
  \left(\SU(4)\ltimes\RR^8\right)\times\RR
\end{equation*}
and, by the results of the appendix, preserves exactly $\frac1{16}$ of 
the supersymmetry.

\sssection{$\left(\Sp(2)\ltimes\RR^8\right)\times\RR$}

Let $K$ again be irreducible with metric given by equation
\eqref{eq:Kmetric} but with holonomy $\Sp(2)$.  Then the warped
product metric given by the expression \eqref{eq:spin7} now has
holonomy
\begin{equation*}
  \left(\Sp(2)\ltimes\RR^8\right)\times\RR
\end{equation*}
and, by the results of the appendix, preserves exactly $\frac3{32}$ of
the supersymmetry.

\sssection{$\left(G_2\ltimes\RR^7\right)\times\RR^2$}

Let $K$ have holonomy $G_2$.  This means that it is reducible.  The
metric is given locally by
\begin{equation*}
  ds^2(K) = g_{mn}(y) dy^m dy^n + (dy^8)^2~,
\end{equation*}
where $m,n$ now only run from $1$ to $7$ and $g_{mn}(y)$, which is
independent of $y^8$, has holonomy $G_2$.  Let $\sigma$ and $\sigma_8$
be arbitrary functions of $x^+$, and consider the following
generalised warped product:
\begin{equation*}
  ds^2 = 2 dx^+ dx^- + a (dx^+)^2 + (dx^1)^2 + e^{2\sigma} g_{mn}
  dy^m dy^n + e^{2\sigma_8} (dy^8)^2~.
\end{equation*}
A similar calculation to the one in Section~\ref{sec:warhol}, which
discusses the degenerate case $\sigma_8=\sigma$, shows that
the holonomy group of the above metric is
\begin{equation*}
  \left(G_2\ltimes\RR^7\right)\times\RR^2~,
\end{equation*}
which, by the results of the appendix, preserves $\frac1{16}$ of the
supersymmetry.

\sssection{$\left(\SU(3)\ltimes\RR^6\right)\times\RR^3$}

Let $K$ now have holonomy $\SU(3)$, so that it is again reducible.
The metric is given locally by
\begin{equation*}
  ds^2(K) = g_{mn}(y) dy^m dy^n + (dy^7)^2 + (dy^8)^2~,
\end{equation*}
where $m,n$ now only run from $1$ to $6$ and $g_{mn}(y)$, which is
independent of $y^7$ and $y^8$, has holonomy $\SU(3)$.  Let $\sigma$,
$\sigma_7$ and $\sigma_8$ be arbitrary functions of $x^+$, and
consider the following generalised warped product:
\begin{equation*}
  ds^2 = 2 dx^+ dx^- + a (dx^+)^2 + (dx^1)^2 + e^{2\sigma} g_{mn}
  dy^m dy^n + \sum_{i=7}^8 e^{2\sigma_i} (dy^i)^2~.
\end{equation*}
A similar calculation to the one in Section~\ref{sec:warhol}, which
again discusses the degenerate case $\sigma_7=\sigma_8=\sigma$, shows
that the holonomy group of the above metric is
\begin{equation*}
  \left(\SU(3)\ltimes\RR^6\right)\times\RR^3~,
\end{equation*}
which, by the results of the appendix, preserves $\frac1{8}$ of the
supersymmetry.

\sssection{$\left(\Sp(1)\ltimes\RR^4\right) \times
\left(\Sp(1)\ltimes\RR^4\right)\times\RR$}

Let $K$ now be reducible with holonomy $\Sp(1)\times\Sp(1)$.  It has
two irreducible components, each having holonomy $\Sp(1)$.  The metric
is given locally by
\begin{equation*}
  ds^2(K) = g^{(1)}_{mn} dy^m dy^n + g^{(2)}_{\Bar m \Bar n} dy^{\Bar
  m} dy^{\Bar n}~,
\end{equation*}
where $m,n$ now only run from $1$ to $4$ and $\Bar m, \Bar n$ run from
$5$ to $8$.  As before $g^{(1)}_{mn}$, which is independent of
$y^5,\dots,y^8$, and $g^{(2)}_{\Bar m \Bar n}$, which is independent
of $y^1,\dots,y^4$, have separately holonomy $\Sp(1)$.  Let $\sigma_1$
and $\sigma_2$ be arbitrary functions of $x^+$, and consider the
following generalised warped product:
\begin{multline*}
  ds^2 = 2 dx^+ dx^- + a (dx^+)^2 + (dx^1)^2\\
  + e^{2\sigma_1} g^{(1)}_{mn} dy^m dy^n + e^{2\sigma_2} g^{(2)}_{\Bar
    m\Bar n} dy^{\Bar m} dy^{\Bar n}~.
\end{multline*}
The similar calculation as in Section~\ref{sec:warhol}, which
again discusses the degenerate case $\sigma_1=\sigma_2=\sigma$, shows
that the holonomy group of the above metric is
\begin{equation*}
  \left(\Sp(1)\ltimes\RR^4\right) \times
  \left(\Sp(1)\ltimes\RR^4\right)\times\RR~,
\end{equation*}
which, by the results of the appendix, preserves $\frac1{8}$ of the
supersymmetry.

\sssection{$\left(\Sp(1)\ltimes\RR^4\right)\times\RR^5$}

Let $K$ have holonomy $\Sp(1)$.  This is a specialisation of the
previous case where $g^{(2)}$ is now the flat metric.  Therefore the
metric is given locally by
\begin{equation*}
  ds^2(K) = g_{mn}(y) dy^m dy^n + dy^{\Bar n} dy^{\Bar n}~,
\end{equation*}
where $m,n$ now only run from $1$ to $4$ and $\Bar m, \Bar n$ run from
$5$ to $8$, and where $g_{mn}$ is independent on $y^5,\dots,y^8$.  Let
$\sigma$, $\sigma_5$, \dots, $\sigma_8$ be arbitrary functions of
$x^+$, and consider the following generalised warped product:
\begin{multline*}
  ds^2 = 2 dx^+ dx^- + a (dx^+)^2 + (dx^1)^2\\
  + e^{2\sigma} g_{mn} dy^m dy^n + \sum_{\Bar n} e^{2\sigma_{\Bar
    n}} dy^{\Bar n} dy^{\Bar n}~.
\end{multline*}
Again as in Section~\ref{sec:warhol}, which discusses the degenerate
case where all the warping factors are equal, one sees that the
holonomy group of the above metric is
\begin{equation*}
  \left(\Sp(1)\ltimes\RR^4\right) \times \RR^5~,
\end{equation*}
which, by the results of the appendix, preserves $\frac1{4}$ of the
supersymmetry.

\sssection{$\RR^9$}

Finally we take $K$ to be flat.  Let $\sigma_m$ for $m=1,\dots,8$ be
arbitrary functions of $x^+$.  The generalised warped product with
metric
\begin{equation*}
  ds^2 = 2 dx^+ dx^- + a (dx^+)^2 + (dx^1)^2 + e^{2\sigma_m}
  dy^m dy^m
\end{equation*}
has holonomy $\RR^9$ for generic $\sigma_m$.  As shown in the appendix
it therefore preserves $\half$ of the supersymmetry.  When
$\sigma_m=0$ for all $m$, we obtain the metric \eqref{eq:richer}.

All the local metrics given in this section are special cases of the
Bryant metric \eqref{eq:bryant} where the family of eight-dimensional
metrics is such that equation \eqref{eq:casd} is satisfied.

\subsection{Ricci-flatness: building the $\M$-wave}
\label{sec:ricciflat}

All the spacetimes resulting from the warping construction just
described admit a parallel null spinor.  This means that provided they
solve the supergravity equations of motion, they are supersymmetric
$\M$-theory vacua.  Because they admit a parallel spinor, they are
Ricci-null.  As explained above, their Ricci tensor has only one
nonzero component $R_{++}$.  Now we investigate whether it is possible
to choose the warp factor $\sigma$ in such a way that the resulting
warped product is Ricci-flat.  We will see that provided that $R_{++}$
depends only on $x^+$, we will be able to choose $\sigma$
appropriately.  Whereas not every supersymmetric Brinkmann wave has
this property, we saw in the Section~\ref{sec:susywaves} that many do
and in the next section we will see one such example in detail.

Let us compute the Ricci tensor of the warped product $M$.  Let
$\Bar\nabla$ denote the Levi-Cività connection of the warped product
$M= B\times_\sigma K$, and $\nabla$ the Levi-Cività connection of each
of the component spaces $B$ and $K$.  The Christoffel symbols can be
read using \eqref{eq:christoffel} from the following:
\begin{align*}
  \Bar\nabla_a \d_b &= \nabla_a \d_b\\
  \Bar\nabla_a \d_m &= \Bar\nabla_m \d_a = \d_a \sigma \, \d_m\\
  \Bar\nabla_m \d_n &= \nabla_m \d_n - g^{ab} \d_b \sigma e^{2\sigma}
  h_{mn} \d_a~.
\end{align*}

The curvature tensor of $M$, similarly adorned with a bar to
distinguish it from that of $B$ and $K$, has the following nonzero
components:
\begin{align*}
  \Bar R_{abc}{}^d &= R_{abc}{}^d\\
  \Bar R_{amb}{}^n &= - \left(\nabla_a \d_b \sigma + \d_a \sigma \d_b
    \sigma \right) \delta_m{}^n\\
  \Bar R_{amn}{}^b &= h_{mn} \nabla_a \left( e^{2\sigma} \d^b
    \sigma \right)\\
  \Bar R_{mnp}{}^q &= R_{mnp}{}^q + e^{2\sigma} g^{ab} \d_a \sigma
  \d_b \sigma \left( h_{np} \delta_m{}^q - h_{mp}
  \delta_n{}^q\right)~.
\end{align*}

The holonomy of the warped product implies that $M$ is Ricci-null,
and this implies that $\Bar R_{++}$ is the only nonzero component in
the Ricci tensor.  But we can verify this directly.  The Ricci
curvature is given by contracting the Riemann curvature:
\begin{align*}
  \Bar R_{ab} &= R_{ab} - (11-d) \left( \nabla_a \d_b \sigma + \d_a
    \sigma \d_b \sigma \right)\\
  \Bar R_{an} &= 0\\
  \Bar R_{mn} &= R_{mn} + h_{mn} \nabla_a \left(e^{2\sigma} \d^a
    \sigma \right) - (10-d) h_{mn} e^{2\sigma} g^{ab}\d_a \sigma \d_b
    \sigma~.
\end{align*}
We expand
\begin{equation*}
  \nabla_a \left( e^{2\sigma} \d^a \sigma \right) = e^{2\sigma} \left( 
  2 \d_a\sigma \d^a \sigma + \d_a \d^a \sigma + \Gamma_{ab}{}^a
  \d^b\sigma \right)~.
\end{equation*}

Indeed, because $\sigma$ only depends on $x^+$, it is harmonic
\begin{equation*}
  \d_a \d^a \sigma = 0~,
\end{equation*}
and its gradient is null
\begin{equation*}
  \d_a\sigma \d^a\sigma = 0~.
\end{equation*}
Similarly the only component of the gradient is $\d^- \sigma$, which
together with the fact that $\Gamma_{a-}{}^b = 0$ (which follows from
the fact that $\d_-$ is parallel) shows that
\begin{equation*}
  \Gamma_{ab}{}^a \d^b\sigma = 0~.
\end{equation*}
Because $K$ is riemannian and admits parallel spinors, it is
Ricci-flat, whence $R_{mn}=0$.  Putting all this together we see that
$\Bar R_{mn} = 0$.  Now because $B$ is a supersymmetric Brinkmann
wave, only $R_{++}$ is nonzero.  Finally, using again that $\sigma$
only depends on $x^+$, we see that the only nonzero component of
\begin{equation*}
  \nabla_a \d_b \sigma + \d_a \sigma \d_b \sigma
\end{equation*}
is the ${}_{++}$ component, which is equal to
\begin{equation*}
  \sigma''  + (\sigma')^2~,
\end{equation*}
where $\sigma' = \d_+\sigma$. In summary, the only nonzero component
of the Ricci tensor of $M$ is given by
\begin{equation*}
  \Bar R_{++} = R_{++} - (11-d) \left(\sigma''  +
  (\sigma')^2\right)~.
\end{equation*}
In terms of $f=e^\sigma$, the vanishing of $\Bar R_{++}$ becomes the
\emph{linear} equation:
\begin{equation}
  \label{eq:ricciflateqn}
  f'' = \frac{1}{11-d} R_{++} \, f~,
\end{equation}
which can be solved for $f$ (and hence $\sigma$) provided that
$R_{++}$ depends \emph{only on $x^+$}.  As was mentioned earlier, this
is not always the case for every supersymmetric Brinkmann wave, but as
we saw in Section~\ref{sec:susywaves} there are plenty of examples for
which this is the case.  In particular, for $d=4$ many examples are
known (see, e.g., \cite{KSMH} and references therein).  A more recent
four-dimensional example is discussed in the next section.

\section{An explicit example: the Nappi--Witten wave}
\label{sec:NW}

The Nappi--Witten background, with the parallelising torsion coming
from the group structure, is a supersymmetric background for
ten-dimensional type II supergravity\footnote{Strictly speaking, the
  vacuum is $N \times K$ with $K$ a suitable six-dimensional manifold,
  e.g., $T^6$, a Calabi--Yau 3-fold,...} and indeed of type II
superstring theory \cite{NW}.  On the other hand, in this section, we
will show how to lift the Nappi--Witten metric to eleven dimensions
and construct from it a supersymmetric vacuum of $d{=}11$
supergravity.

\subsection{The Nappi--Witten geometry}

We start by discussing the geometry of the Nappi--Witten metric,
paying close attention to its holonomy.

The Nappi--Witten spacetime \cite{NW} describes a four-dimensional
solvable Lie group $N$ possessing a bi-invariant lorentzian metric.
The Lie algebra $\fn$ is the universal central extension of the
two-dimensional euclidean algebra, with generators $\{X_1, X_2, X_-,
X_+\}$ obeying the following non-vanishing Lie brackets:
\begin{equation}
  \label{eq:ISO(2)c}
  [X_i,X_j] = \epsilon_{ij} X_- \qquad\text{and}\qquad [X_+,X_i] =
  \epsilon_{ij} X_j~,
\end{equation}
where $i,j=1,2$ and $\epsilon_{12} = -\epsilon_{21} = 1$.  Up to
scale, there is a one-parameter family of invariant scalar products:
\begin{equation}
  \label{eq:gmetric}
  \langle X_i, X_j \rangle = \delta_{ij}~,\quad
  \langle X_+, X_+ \rangle = b\quad\text{and}\quad
  \langle X_+, X_- \rangle = 1~.
\end{equation}
Clearly these metrics are non-degenerate, ad-invariant, and
lorentzian.

The parameter $b$ is inessential.  Changing basis to $\{X'_1, X'_2,
X'_+,X'_-\}$ where $X'_i=X_i$ for $i=1,2$, $X'_- = X_-$ and $X'_+ =
X_+ - \half b X_-$, the metric now has $b=0$.  Since $X_-$ is central
and $X_+ \not\in [\fn,\fn]$, this change of basis is an automorphism
of $\fn$.  Therefore without loss of generality we will set $b=0$ and
drop the primes in the generators.

A convenient coordinate chart for $N$ is given by
\begin{equation}
  \label{eq:chart}
  n(x) = e^{x^1 X_1 + x^2 X_2}\, e^{x^-X_- + x^+ X_+} \in N~,
\end{equation}
where the coordinates $(x^1,x^2,x^-,x^+)$ take values in $\RR^4$.
Strictly speaking, this is the universal cover of the Nappi--Witten
spacetime.  It is possible to consider quotients where the coordinate
$x^+$ is periodic, but we will not do so here, since periodicity would
spoil the construction in Section~\ref{sec:d=11} of a supersymmetric
background for $d{=}11$ supergravity.  We will comment on this again
later on.

The left- and right-invariant Maurer--Cartan forms are given by
\begin{equation}
  \label{eq:MaurerCartan}
  \btheta_L = n(x)^{-1} dn(x) \quad\text{and}\quad
  \btheta_R = dn(x)\, n(x)^{-1}~.
\end{equation}
These are one-forms on $N$ taking values in $\fn$.  We can define a
left-invariant metric $ds^2$ in $N$ by
\begin{equation}
  \label{eq:Gmetric}
  ds^2 = \langle \btheta_L, \btheta_L \rangle \equiv \btheta_L^i
  \btheta_L^i + 2 \btheta_L^+ \btheta_L^-~,
\end{equation}
with summation over $i=1,2$ implied.  Because the scalar product
\eqref{eq:gmetric} is ad-invariant, we can also write $ds^2$ in terms 
of the right-invariant Maurer--Cartan forms as
\begin{equation*}
  ds^2 = \langle \btheta_R, \btheta_R \rangle~,
\end{equation*}
which shows that $ds^2$ is also right-invariant.

In terms of the local coordinate chart \eqref{eq:chart}, the metric
\eqref{eq:Gmetric} becomes
\begin{equation}
  \label{eq:metric}
  ds^2 = 2 dx^+ dx^- + \epsilon_{ij} x^j dx^i dx^+ + dx^i dx^i~,
\end{equation}
which is of the form \eqref{eq:susybrinkmann4d} with $a=0$ and $b=1$.
Hence from our previous analysis we know that this is a supersymmetric
Brinkmann wave.  Nevertheless let us compute some of the geometrical
objects associated with this metric.

The non-vanishing coefficients of the Levi-Cività connection are
\begin{equation}
  \label{eq:connection}
  \nabla_+ \d_i = \nabla_i \d_+ = -\tfrac14 x^i \d_- - \half
  \epsilon_{ij} \d_j~,
\end{equation}
where $i,j=1,2$.  The nonzero components of the Riemann curvature
tensor are given by
\begin{equation}
  \label{eq:curvature}
  R_{+i+}{}^j = \tfrac14 \delta_{ij} \quad\text{and}\quad
  R_{+i+}{}^- = -\tfrac18 \epsilon_{ij} x^j~,
\end{equation}
from where we compute the Ricci tensor, whose only nonzero component
is
\begin{equation}
  \label{eq:ricci}
  R_{++} = \half~.
\end{equation}
Hence the Nappi--Witten geometry is \emph{not} Ricci-flat, although
the curvature scalar does vanish.

In order to construct the spin connection, we first need to choose a
pseudo-orthonormal coframe.  It will prove convenient to use yet a
third coframe $\btheta$, different from the either of the two
Maurer--Cartan coframes.  Let us define
\begin{equation}
  \label{eq:coframe}
  \btheta^{\Hat i} = dx^i + \half \epsilon_{ij} x^j dx^+~,\quad
  \btheta^{\Hat -} = dx^- - \tfrac18 |x|^2 dx^+\quad\text{and}\quad
  \btheta^{\Hat +} = dx^+~,
\end{equation}
where $|x|^2 \equiv x^i x^i$.  In terms of this coframe, the metric
\eqref{eq:metric} is simply
\begin{equation}
  \label{eq:minkowski}
  ds^2 = \btheta^{\Hat i} \btheta^{\Hat i} + 2 \btheta^{\Hat +}
  \btheta^{\Hat -}~.
\end{equation}
The connection one-form $\bomega$ is defined by the first structure
equation \eqref{eq:fse}.  Defining $\bomega^{\Hat a\Hat b} = 
-\eta^{\Hat b \Hat c} \bomega_{\Hat c}{}^{\Hat a}$, we can express the
nonzero components of the connection one-form as follows:
\begin{equation}
  \label{eq:spinconn}
  \omega_+^{\Hat 1\Hat 2} = -\half~,\quad
  \omega_+^{\Hat 1\Hat -} = -\tfrac14 x^1\quad\text{and}\quad
  \omega_+^{\Hat 2\Hat -} = -\tfrac14 x^2~.
\end{equation}

The curvature two-form $\bOmega$ follows from the second structure
equation \eqref{eq:sse}.  It follows from equation \eqref{eq:spinconn}
that the quadratic term does not contribute, from where we see that
the only nonzero components of the curvature two-form are
\begin{equation}
  \label{eq:curv2formcomps}
  \Omega_{1+}{}^{\Hat 1\Hat -} = -\tfrac14 \quad\text{and}\quad
  \Omega_{2+}{}^{\Hat 2\Hat -} = -\tfrac14~,
\end{equation}
which can be shown to agree with \eqref{eq:curvature}.

By the Ambrose--Singer theorem, the holonomy algebra is the Lie
subalgebra $\fh \subset \so(3,1)$ spanned by the $\half
\Omega_{ab}{}^{\Hat c \Hat d} e_{\Hat c} \wedge e_{\Hat d}$ for all
$a,b$, where we are identifying the Lie algebra $\so_{3,1}$ of the
Lorentz group with the bivectors $\bigwedge^2 \MM^4$ in Minkowski
spacetime.  In the case at hand the holonomy algebra is the
two-dimensional abelian subalgebra generated by $h_i \equiv e_{\Hat
i}\wedge e_{\Hat -}$ for $i=1,2$.   If $\{e_{\Hat a}\}$ is a
pseudo-orthonormal basis for $\MM^4$ with inner product $\langle
e_{\Hat a}, e_{\Hat b}\rangle = \eta_{\Hat a\Hat b}$, then the
infinitesimal Lorentz transformations are given by:
\begin{equation}
  \label{eq:affine}
  [e_{\Hat a}\wedge e_{\Hat b}, e_{\Hat c}] = \eta_{\Hat a\Hat c}
  e_{\Hat b} - \eta_{\Hat b\Hat c} e_{\Hat a}~.
\end{equation}

Notice that the action of the holonomy representation in $\MM^4$ is
reducible, yet it is indecomposable.  To see this let us consider the
action of the $h_i$ on the basis $\{e_{\Hat a}\}$ of $\MM^4$.  It acts 
via null rotations:
\begin{gather*}
  [h_1, e_{\Hat 1}] = e_{\Hat -}\quad [h_1,e_{\Hat 2}] = 0 \quad
  [h_1,e_{\Hat +}] = -e_{\Hat 1} \quad
  [h_1,e_{\Hat -}] = 0\\
  [h_2, e_{\Hat 1}] = 0 \quad [h_2,e_{\Hat 2}] = e_{\Hat -} \quad
  [h_2,e_{\Hat +}] = -e_{\Hat 2} \quad [h_2,e_{\Hat -}] = 0~.
\end{gather*}
It is clear that the line spanned by $e_{\Hat -}$ is a
subrepresentation which does not split.  The $\fh$-invariant forms are
given in Table~\ref{tab:invariants}.  By the holonomy principle these
invariants give rise to parallel forms in the Nappi--Witten
spacetime.

\begin{table}[h!]
\renewcommand{\arraystretch}{1.1}
\begin{tabular}{|>{$}c<{$}|>{$}c<{$}|}\hline
  \text{Rank} & \text{Parallel Forms}\\ \hline\hline
  0 & 1\\
  1 & \btheta^{\Hat +} = dx^+\\
  2 & \btheta^{\Hat i} \wedge \btheta^{\Hat +} = dx^i \wedge dx^+ \\
  3 & \btheta^{\Hat 1} \wedge \btheta^{\Hat 2} \wedge \btheta^{\Hat +}
  = dx^1 \wedge dx^2 \wedge dx^+\\
  4 & \btheta^{\Hat 1} \wedge \btheta^{\Hat 2} \wedge \btheta^{\Hat -}
  \wedge \btheta^{\Hat +} = dx^1 \wedge dx^2 \wedge dx^- \wedge dx^+\\
  \hline
\end{tabular}
\vspace{8pt}
\caption{Parallel forms in the Nappi--Witten spacetime}
\label{tab:invariants}
\end{table}

The Nappi--Witten spacetime $N$ is parallelisable (it is a Lie group)
and hence it admits a spin structure; although, as the curvature
calculation shows, the parallelising connection is not the Levi-Cività
connection.  Nevertheless, as we will see presently, the Nappi--Witten
manifold admits parallel spinors.

We can first of all detect the existence of parallel spinors by
studying the action of the holonomy algebra on the spinors.  The
spinors are in a representation of $\Spin(3,1) \cong \SL(2,\CC)$,
which sits inside the Clifford algebra $\Cl(3,1)$.  Let $\varrho:\fh
\to \Cl(3,1)$ denote the spinorial representation of the holonomy
algebra, defined by
\begin{equation*}
  \varrho(h_i) = \half \Gamma_{\Hat i}\Gamma_{\Hat -}\quad\text{for
  $i=1,2$,}
\end{equation*}
where the $\{\Gamma_{\Hat a}\} = \{\Gamma_{\Hat i},\Gamma_{\Hat
-},\Gamma_{\Hat +}\}$ satisfy
\begin{equation}
  \label{eq:Cl(3,1)}
  \{\Gamma_{\Hat a},\Gamma_{\Hat b}\} = - 2 \eta_{\Hat a\Hat b} \1~.
\end{equation}
By the holonomy principle, parallel spinors will be in one-to-one
correspondence with $\fh$-invariant spinors.  A spinor $\varepsilon$
is $\fh$-invariant if and only if
\begin{equation*}
  (\forall i)\quad \Gamma_{\Hat i} \Gamma_{\Hat -} \varepsilon = 0
  \quad\iff\quad \Gamma_{\Hat -} \varepsilon = 0~.
\end{equation*}
Since $\Gamma_{\Hat -}^2 = 0$ and $\{\Gamma_{\Hat -},\Gamma_{\Hat +}\}
= -2 \1$, it follows that
\begin{equation*}
  \Gamma_{\Hat -} \varepsilon = 0 \quad\iff \quad \varepsilon =
  \Gamma_{\Hat -} \chi~,
\end{equation*}
where $\chi = -\half \Gamma_{\Hat +}\varepsilon$.  In other words,
one half of the spinors are parallel.

We can solve explicitly for the parallel spinors as follows.  Let
$\varepsilon$ be a parallel spinor.  It satisfies the differential
equation
\begin{equation*}
  \d_a \varepsilon = -\tfrac14 \omega_a{}^{\Hat c\Hat d} \Gamma_{\Hat
  c\Hat d} \varepsilon~.
\end{equation*}
From the explicit expression of the spin connection
\eqref{eq:spinconn}, this equation becomes
\begin{equation*}
  \d_i \varepsilon = \d_-\varepsilon = 0 \quad\text{and}\quad
  \d_+ \varepsilon = \tfrac14 \Gamma_{\Hat 1\Hat 2} \varepsilon~,
\end{equation*}
where we have used that $\Gamma_{\Hat -} \varepsilon = 0$.  Since
$\Gamma_{\Hat 1\Hat 2}^2 = -\1$, it is a complex structure.  Since it
commutes with $\Gamma_{\Hat -}$ it preserves those spinors in its
kernel and can be diagonalised there.  Therefore let us write
$\varepsilon = \varepsilon_+ + \varepsilon_-$, where $\Gamma_{\Hat
  1\Hat 2} \varepsilon_\pm = \pm i \varepsilon_\pm$, and $\Gamma_{\Hat
  -} \varepsilon_\pm = 0$.  Then
\begin{equation*}
  \d_+ \varepsilon_\pm = \pm \tfrac{i}{4} \varepsilon_\pm~,
\end{equation*}
which has solutions
\begin{equation*}
  \varepsilon = \chi_+ e^{i x^+/4} + \chi_- e^{-i x^+/4}~,
\end{equation*}
where $\chi_\pm$ are constant spinors satisfying $\Gamma_{\Hat -}
\chi_\pm = 0$ and $\Gamma_{\Hat 1\Hat 2} \chi_\pm = \pm i \chi_\pm$.
We can further impose a reality condition on $\varepsilon$ by taking
$\chi_+ = \chi_-^*$.  Notice that since spinors can change by a sign
around a non-contractible loop, it would be possible to periodically
identify $x^+ \sim x^+ + 4\pi N$, for some integer $N$, and still have
parallel spinors.

\subsection{Lifting the Nappi--Witten background}
\label{sec:d=11}

We now discuss how to lift the Nappi--Witten background to eleven
dimensions while preserving supersymmetry.

Let $K$ be a seven-dimensional riemannian manifold with holonomy group 
contained in $G_2 \subset \SO(7)$.  Let $\sigma$ be an arbitrary
function of $x^+$ and let us consider the warped product $M = N
\times_\sigma K$ with metric
\begin{equation}
  \label{eq:warped}
  ds^2 = ds^2(N) + e^{2\sigma}\, ds^2(K)~.
\end{equation}

From the results in Section~\ref{sec:lift} we know that the above
warped product has holonomy contained in $\left(G_2 \ltimes
\RR^7\right) \times \RR^2$ and therefore it admits a parallel null
spinor.  The spacetime will in addition be Ricci-flat provided that
the function $f = e^\sigma$ obeys the differential equation
\eqref{eq:ricciflateqn}, which using equation \eqref{eq:ricci} becomes
\begin{equation*}
  f'' = \frac{1}{14} f = 0~.
\end{equation*}
Notice that the solutions to this equation grow asymptotically for
$|x^+| \to \infty$, and hence we are not allowed to periodically
identify $x^+$, and are forced instead to work on the universal cover
of the Nappi--Witten spacetime.

If the metric on $K$ has holonomy precisely $G_2$ then $M$ admits a
total of $2$ parallel spinors and this means that the spacetime
preserves $\frac1{16}$ of the supersymmetry.  At the other extreme, if
$K$ is flat, then the holonomy group of $M$ is $\RR^9$ and it admits a
total of $16$ parallel spinors, accounting for $\half$ of the
supersymmetry.  By choosing the holonomy group of $K$ to be any
admissible group in between: $\SU(3)$ and $\Sp(1)=\SU(2)$, we can
obtain $\frac18$ and $\frac14$ the supersymmetry, respectively.  In
other words, the spacetime preserves $\half$ of the supersymmetry
preserved by the vacuum $\MM^4 \times K$.

\section{Some solutions with $F\neq 0$}
\label{sec:addflux}

In this section we will investigate the possibility of relaxing the
assumption of vanishing four-form while preserving the form of the
metric.  As we have seen above, the metrics we constructed depend on
some arbitrary functions which are then further constrained to make
the spacetime Ricci-flat, as demanded by the equations of motion.
With a nonzero four-form, the equations of motion will change: in
particular the metric will not be Ricci-flat; but this just changes
the conditions on the arbitrary functions the metric depends on.
Therefore it seems \emph{a priori} that there might be room for
introducing a non-zero $F$ while preserving the form of the metric.

Indeed, as we will now see, it will be possible in some cases to add a
nonzero four-form while both preserving supersymmetry and satisfying
the Maxwell and Einstein equations of motion.  Our approach derives
its inspiration from the work of Hull \cite{Mwave}, who took the
metric given by equation \eqref{eq:richer} and showed how to add
nonzero gravitino and four-form to obtain a supersymmetric vacuum of
eleven-dimensional supergravity.  In fact, Hull's solution can be
specialised by setting the gravitino to zero and hence one obtains in
this way a supersymmetric bosonic vacuum with nonzero four-form.

The metric \eqref{eq:richer} is rather special, since it has holonomy
$\RR^9$.  In this section we will investigate whether it is possible
to make supersymmetric vacua with nonzero $F$ and with metrics of
larger holonomy.  We will see that this will be possible for metrics
with all the holonomy groups in Table~\ref{tab:mwaves} \emph{except}
for those with the largest possible holonomy $G$ in
\eqref{eq:exotic}.

\subsection{Supersymmetric bosonic vacua with nonzero flux}

We now go about adding $F$ to the background defined by Bryant's
metric \eqref{eq:bryant}, in such a way that the equations of motion
are satisfied.  We first need to write down a suitable $F$.  Because
the Bryant metric admits a parallel null spinor it is Ricci-null, and
because $\d_-$ is parallel, this means that the only nonzero component
of the Ricci tensor is $R_{++}$.  This implies that the scalar
curvature $R$ vanishes and if the equations of motion \eqref{eq:genEM}
are to be satisfied, then the only nonzero component of the
energy-momentum tensor is $T_{++}$.  This means that the four-form $F$
is null; in fact, $F$ must be of the form
\begin{equation}
  \label{eq:ansatzF}
  F = dx^+ \wedge \Phi~,
\end{equation}
where $\Phi$ is a $3$-form with components $\Phi_{\mu\nu\rho}$ where
$\mu,\nu,\rho$ take the values $1$ to $9$; in other words, $\d_-
\lrcorner \Phi = 0$.  We now impose the Bianchi identity and the
equation of motion for $F$.

The Bianchi identity ($dF=0$) is satisfied provided that $\Phi$ is
closed as a $3$-form in the $9$-dimensional space $X_9$ parametrised
by $x^\mu$ for $\mu=1,\dots,9$, and that it is independent of $x^-$.
The dependence on $x^+$ is not constrained.

The equation of motion \eqref{eq:feom} simply says that $F$ is
co-closed: $d\star F = 0$.  The Hodge dual of $F$ relative to the
Bryant metric is given by 
\begin{equation*}
  \star F = \half dx^+ \wedge \star_9 \Phi~,
\end{equation*}
where $\star_9$ is the Hodge dual in $X_9$ relative to the
$x^+$-dependent metric
\begin{equation}
  \label{eq:metricX9}
  ds^2(X_9) = (dx^9)^2 + g_{ij}dx^i dx^j~.
\end{equation}
Therefore the equations of motion simply says that $\star_9 \Phi = 0$
is a closed form on $X_9$.  In other words, $\Phi$ is an
$x^+$-dependent family of harmonic forms on $X_9$ (relative to the
$x^+$-dependent family of metrics \eqref{eq:metricX9}).  In order to
constraint $\Phi$ further it will be necessary to impose that the
background be supersymmetric.

The condition for a bosonic background to be supersymmetric is the
vanishing of the supersymmetry variation of the gravitino, which is
given by equation \eqref{eq:gravshift}, with $\btheta_a(F)$ given by
\eqref{eq:cliffconn}.  In order to find solutions to this equation
we will make the following \emph{assumption}: that the spinors
$\varepsilon$ satisfy
\begin{equation}
  \label{eq:assumption}
  \Gamma_{\Hat -} \varepsilon = 0~.
\end{equation}
It might be possible to prove that this assumption is actually forced
upon us, by a full analysis of the integrability of equation
\eqref{eq:gravshift}, but thus far we have been unable to prove it.
At any rate, together with the form \eqref{eq:ansatzF} for $F$, this
assumption implies that $\theta_a \cdot \varepsilon = 0$ for all $a$
except $a=+$, for which
\begin{equation}
  \label{eq:cliffconn+}
  \theta_+(F) \cdot \varepsilon = -\tfrac16 \Phi \cdot \varepsilon~,
\end{equation}
where $\Phi \cdot \varepsilon = \frac16
\Phi_{\mu\nu\rho}\Gamma^{\mu\nu\rho} \epsilon$.  Equation
\eqref{eq:gravshift} now becomes
\begin{equation}
  \label{eq:killing}
  \nabla_- \varepsilon = 0~,\qquad
  \nabla_\mu \varepsilon = 0 \qquad \text{and} \qquad
  \nabla_+ \varepsilon = -\tfrac{1}{6} \Phi \cdot \varepsilon~.
\end{equation}
The integrability of these equations implies that $\Phi$ is a parallel
form (compare with \cite{Hull-planewaves}).  In other words, the
metric \eqref{eq:metricX9} on $X_9$ should admit a parallel $3$-form.

Let us now show that the third equation in \eqref{eq:killing} is
actually just
\begin{equation}
  \label{eq:+killing}
  \d_+ \varepsilon = -\tfrac{1}{6} \Phi \cdot \varepsilon~,
\end{equation}
which basically fixes the dependence of $\varepsilon$ on $x^+$.  To
prove this, all we need to show is that $\omega_+^{\Hat a\Hat b}
\Gamma_{\Hat a \Hat b} \varepsilon = 0$.  This follows from the
condition that $\Gamma_{\Hat -}\varepsilon = 0$ and the fact that the
variation of the metric is conformal anti-self dual.

To see this let us write down the spin connection for the Bryant
metric \eqref{eq:bryant}.  Consider the following pseudo-orthonormal
coframe:
\begin{equation*}
  \btheta^{\Hat +} = dx^+ \qquad \btheta^{\Hat -} = dx^- + \half a
  dx^+ \qquad \btheta^{\Hat 9} = dx^9 \qquad \btheta^{\Hat i}~,
\end{equation*}
where $\btheta^{\Hat i}$ is an $x^+$-dependent orthonormal coframe for
the metric $g_{ij}$.  The connection one-form is given by the first
structure equation \eqref{eq:fse}.  Because $\btheta^{\Hat i}$ depend
on $x^+$ we need to introduce a matrix $V_{\Hat i \Hat j}$ to
represent the variation of the coframe with respect to $x^+$.
According to \cite{Bryant-spinors} for a conformal anti-self dual
variation, the matrix $V_{\Hat i \Hat j}$ is symmetric: $V_{\Hat i\Hat
j} = V_{\Hat j \Hat i}$.  In this case, the nonzero components of the
connection one-form are given by
\begin{align*}
  \bomega_{\Hat i}{}^{\Hat -} &= \half \d_{\Hat i} a \, \btheta^{\Hat +} - 
  V_{\Hat i \Hat j} \, \btheta^{\Hat j}\\
  \bomega_{\Hat 9}{}^{\Hat -} &= \half \d_{\Hat 9} a\, \btheta^{\Hat
  +}~,
\end{align*}
together with the $x^+$-dependent connection one-form $\bomega_{\Hat
  i}{}^{\Hat j}$ of the metric $g_{ij}$.  The nonzero components of
$\omega_+{}^{\Hat a \Hat b}$ are therefore $\omega_+{}^{\Hat i \Hat
  -}$ and $\omega_+{}^{\Hat 9 \Hat -}$.  Since the spinor
$\varepsilon$ obeys $\Gamma_{\Hat -} \varepsilon = 0$, it follows that
$\omega_+{}^{\Hat a \Hat b} \Gamma_{\Hat a\Hat b} \varepsilon = 0$,
whence $\nabla_+ \varepsilon = \d_+\varepsilon$, as claimed. 

If the variation does not obey \eqref{eq:casd}, so that the matrix $V$
has an skew-symmetric part, then the ${\Hat i\Hat j}$ components of
the connection one-form would be modified by the addition of $V_{[\Hat 
i \Hat j]} \, \btheta^{\Hat +}$.  Therefore there would be another
nonzero component $\omega_+{}^{\Hat i \Hat j}$, which would spoil the
argument.

In summary, $\varepsilon$ is an $x^+$-dependent doublet of parallel
spinors on $X_9$ satisfying equation \eqref{eq:+killing}, and the
condition $\Gamma_{\Hat -} \varepsilon = 0$ chooses one of the spinors
in the doublet.  The number of supersymmetries which this background
preserves is then equal to the number of parallel spinors in $X_9$.
In other words, the same fraction $\nu$ of supersymmetry is preserved
as in the case of vanishing $F$.

Finally, the Einstein equations of motion read 
\begin{equation}
  \label{eq:geom}
  R_{++} = \tfrac16 \Phi_{\mu\nu\rho} \Phi^{\mu\nu\rho}~.
\end{equation}
Because $\Phi$ is parallel in $X_9$, the right-hand side of this
equation is actually only a function of $x^+$.  This equation can be
used to constrain the function $a$.

In summary, the Bryant metric \eqref{eq:bryant} with $F$ given by
\eqref{eq:ansatzF}, with $\Phi$ an $x^+$-dependent parallel $3$-form
relative to the $x^+$-dependent metric \eqref{eq:metricX9}, is a
supersymmetric bosonic vacuum of eleven-dimensional supergravity
provided that the arbitrary function in the Einstein metric $a$ is
such that equation \eqref{eq:geom} is satisfied.  This background
preserves the same amount of supersymmetry as the purely gravitational
background.  Notice that the existence of a parallel $3$-form rules
out only one of the possible holonomy groups of the metric: the
maximal one $G$ in \eqref{eq:exotic}, since there are no parallel
$3$-forms on a nine-dimensional manifold $X_9$ with holonomy exactly
$\Spin(7)$.  For any of the other groups in Table~\ref{tab:mwaves}
there is always a parallel $3$-form and hence supersymmetric bosonic
vacua with $F\neq 0$ exist and preserve the fraction $\nu$ of the
supersymmetry listed in that table.

\subsection{Adding a gravitino}
\label{sec:addgrav}

It is also possible to give a vacuum expectation value to the
gravitino.  In fact, as was done in \cite{Mwave} for the metric
\eqref{eq:richer}, one can try the Ansatz \cite{Urrutia}
\begin{equation}
  \label{eq:urrutia}
  \psi_+ = \chi\quad\text{and}\quad
  \psi_a = 0\quad\text{for $a\neq +$.}
\end{equation}
In this Ansatz both the gravitino torsion and the gravitino
energy-momentum tensor vanish.  It also leads to trivial
supercovariantisation of the spin connection and the four-form $F$.
Therefore the equations of motion of $F$ and of the metric are not
altered, and neither is the supersymmetry variation of the gravitino.
However we now have that the bosonic fields do transform under
supersymmetry and that the gravitino is subject to the
Rarita--Schwinger equation.  Requiring supersymmetry demands that
$\chi$ be a parallel spinor in $X_9$, whereas the Rarita--Schwinger
equation fixes the dependence of this spinor on $x^+$.

\section{Conclusions and open problems}
\label{sec:final}

In this paper we have initiated a systematic analysis of possible
supersymmetric bosonic $\M$-theory vacua with zero four-form.  As we
have seen there are two main types of vacua: static vacua which
generalise the Kaluza--Klein monopole and non-static vacua
generalising the $\M$-wave.  We have also seen how to obtain from
these vacua other supersymmetric vacua with non-vanishing four-form.

A convenient invariant, which refines the more standard supersymmetry
fraction, is the holonomy group of the metric defining the vacuum.
The holonomy groups of the static vacua are the well-known riemannian
holonomy groups admitting parallel spinors: they are given in
Table~\ref{tab:static}.  In contrast, the holonomy groups for the
supersymmetric $\M$-waves are lorentzian holonomy groups which have
not been studied much.  The indecomposable groups are listed in
Table~\ref{tab:mwaves}.  The rest of the paper has been mostly devoted
to the construction of supersymmetric $\M$-waves with the holonomy
groups in Table~\ref{tab:mwaves}.  The construction consists of warped
products of lower dimensional supersymmetric (Brinkmann) $pp$-waves
with riemannian manifolds admitting parallel spinors.  In particular,
all groups in the table can be obtained as holonomy groups of warped
products of three-dimensional waves with eight-dimensional manifolds
with holonomy contained in $\Spin(7)$.  The fractions of supersymmetry
preserved by these vacua (both static and non-static) belong to the
set $\{1, \half, \frac14, \frac3{16}, \frac18, \frac3{32}, \frac1{16},
\frac1{32}\}$.  It is thus a natural question to ask whether some of
these vacua are actually dual to vacua consisting of intersecting
branes, where some of these same fractions occur.

Many of the metrics described here have spacelike isometries, along
which the metric can be reduced dimensionally to a supersymmetric wave
of type IIA string theory, of the type that has been much studied in
the literature (see, for example, \cite{BKO,HoTs,GSR}).  Many of these
supersymmetric waves belong to a duality multiplet, which may contain
other type IIA vacua which can be oxidised back to eleven dimensions.
In this way we can construct dual vacua to the $\M$-waves described in
this paper.  We believe that it is an interesting problem to
investigate these dual vacua.  Similarly, every supersymmetric
$\M$-wave possesses a null isometry.  It would be very interesting to
perform a null reduction \cite{JuliaNicolai} of these vacua and then
try to dualise and oxidise back to eleven dimensions.

Finally, all the metrics we have discussed in this paper are
\emph{local}.  We have not analysed the global (spatial or causal)
structure of the spacetimes admitting these metrics.  This is an
interesting area for further research.

\section*{Acknowledgements}

Sonia Stanciu contributed in the early stages of this work and it is
my pleasure to thank her for that.  It is also a pleasure to thank
Chris Hull, Tomás Ortín and Arkady Tseytlin for useful discussions,
and especially Robert Bryant for sending me his notes on spinors and
for much insightful correspondence.

\appendix

\section{Some exotic lorentzian holonomy groups}

In this appendix we collect some facts about the group $G$ in equation
\eqref{eq:exotic} and some of its subgroups.  More precisely we
describe the Lie algebras and how they are embedded in the lorentzian
spin algebra.  As usual when discussing the representations of a spin
algebra it is convenient to work with the Clifford algebra into which
the spin algebra is embedded.  The eleven-dimensional spin algebra
$\so(10,1)$ is naturally embedded inside the Clifford algebras
$\Cl(10,1)$ and $\Cl(1,10)$.  We choose to work with the latter
algebra.  Our Clifford algebra conventions are as in \cite{LM}.  In
particular, the defining relations for $\Cl(1,10)$ are
\begin{equation*}
  \{\Gamma_A, \Gamma_B \} = 2\eta_{AB} \1~,
\end{equation*}
where $\eta_{AB}$ is mostly plus.  Since we want to be as explicit as
possible, our first task is to find an explicit realisation of
$\Cl(1,10)$.

\subsection{An explicit real realisation of $\Cl(1,10)$}

As a real associative algebra, $\Cl(1,10)$ is isomorphic to two copies 
of the algebra of real $32\times 32$ matrices:
\begin{equation*}
\Cl(1,10) \cong \Mat(32,\RR)\oplus\Mat(32,\RR)~.
\end{equation*}
This means that there are precisely two inequivalent irreducible
representations: both real and of dimension $32$.  Of course both
representations are equivalent under the spin algebra $\so(10,1)$, and
isomorphic to the spinor representation $\Delta$.  Choosing a set of
generators $\Gamma_0, \Gamma_1, \ldots, \Gamma_9, \Gamma_\natural$ for
$\Cl(1,10)$, their product $\Gamma_{012\cdots9\natural}$ commutes with
all $\Gamma_A$ and squares to one.  Hence by Schur's lemma it is
$\pm\1$ on an irreducible representation.  We will identify $\Delta$
with the irreducible representation of the Clifford algebra for which
$\Gamma_{012\cdots9\natural}$ takes the value $-\1$.  This means that
$\Gamma_0 = \Gamma_{1\cdots9\natural}$, where $\Gamma_1, \ldots,
\Gamma_9, \Gamma_\natural$ generate the ten-dimensional Clifford
algebra $\Cl(0,10)$.  The Clifford algebra $\Cl(0,10)$ is isomorphic
to $\Cl(8)\otimes\Cl(0,2)$, where the isomorphism is given explicitly
as follows in terms of generators.  Let $\Gamma'_i$ for
$i=1,2,\ldots,8$ denote the generators for $\Cl(8)$ and let
$\Gamma''_1$ and $\Gamma''_2$ denote the generators for $\Cl(0,2)$.

The $\Gamma'_i$ can be constructed explicitly in terms of the
octonions $\OO$.  The construction of the two irreducible
representations of $\Cl(7)$ in terms of octonions is well known: see,
for example, \cite{LM}.  Let $\{o_i\}$, $i=1,\ldots,7$, be a set of
imaginary octonion units.  Then left multiplication $L_i$ and right
multiplication $R_i$ by $o_i$ on $\OO$ define the two inequivalent
irreducible representations of the Clifford algebra $\Cl(7)$.  Either
representation can be used in order to build the unique irreducible
representation of $\Cl(8)$---we choose $L_i$.  Thus we define
\begin{equation*}
\Gamma'_i = \begin{pmatrix}
            0 & L_i \\ L_i & 0
            \end{pmatrix}
\quad\text{for $i=1,\ldots,7$; and}\quad
\Gamma'_8 = \begin{pmatrix}
            0 & -\1 \\ \1 & 0
            \end{pmatrix}~,
\end{equation*}
yielding a manifestly real $16$-dimensional representation of $\Cl(8)
\cong \Mat(16,\RR)$.

As associative algebras, $\Cl(0,2) \cong \Mat(2,\RR)$, so we can
choose a basis
\begin{xalignat*}{2}
\Gamma''_1 &= \sigma_1 = \begin{pmatrix}0& 1\\ 1 &
0\end{pmatrix}\qquad & \Gamma''_2 &= \sigma_3 = \begin{pmatrix}1& 0\\
0 & -1\end{pmatrix}\\
\therefore\qquad \Gamma''_3 & = \Gamma''_1\Gamma''_2 = -i\sigma_2 =
\begin{pmatrix}0& -1\\ 1 & 0\end{pmatrix}~.
\end{xalignat*}
Then the generators of $\Cl(0,10)$ are given by
\begin{xalignat*}{2}
\Gamma_0 &= \Gamma'_9 \otimes \Gamma''_3 \qquad &
\Gamma_i &= \Gamma'_i \otimes \Gamma''_3\qquad\text{for
$i=1,2,\ldots,8$}\\
\Gamma_9 &= \1 \otimes \Gamma''_2 \qquad & \Gamma_\natural &= \1
\otimes \Gamma''_1~,
\end{xalignat*}
where $\Gamma'_9 \equiv \Gamma'_1\Gamma'_2\cdots\Gamma'_8$.  This
decomposition induces an isomorphism $\Delta \cong \RR^{16} \otimes
\RR^2$, so that we can write the eleven-dimensional spinors as
two-component objects, each component being a sixteen-dimensional real
spinor of $\Cl(8)$. Therefore in terms of $\Cl(8)$ generators we have
\begin{alignat}{2}\label{eq:gammas}
\Gamma_0 &= \begin{pmatrix}0 & -\Gamma'_9 \\ \Gamma'_9 &
0\end{pmatrix} \qquad & \Gamma_i &=
\begin{pmatrix}0 & -\Gamma'_i\\ \Gamma'_i &
0\end{pmatrix}\quad\text{for $i=1,2,\ldots,8$}\notag\\
\Gamma_9 &= \begin{pmatrix}\1 & 0\\ 0 & -\1\end{pmatrix} \qquad
& \Gamma_\natural &= \begin{pmatrix}0 & \1 \\ \1 & 0\end{pmatrix}~.
\end{alignat}

The standard basis for the Lie algebra $\so(10,1) \subset \Cl(1,10)$
is given by $\Sigma_{AB} = - \half \Gamma_{AB}$, which in the chosen
realisation becomes
\begin{alignat*}{2}
\Sigma_{ij} &= \begin{pmatrix}\Sigma'_{ij} & 0 \\ 0 &
\Sigma'_{ij}\end{pmatrix}\qquad & \Sigma_{i\natural} &=
\begin{pmatrix}\half \Gamma'_i & 0 \\ 0 & -\half
\Gamma'_i\end{pmatrix}\\
\Sigma_{i9} &= \begin{pmatrix}0 & -\half\Gamma'_i\\ -\half
\Gamma'_i & 0\end{pmatrix}\qquad & \Sigma_{0i} &=
\begin{pmatrix}\half \Gamma'_9\Gamma'_i & 0 \\ 0 & \half
\Gamma'_9\Gamma'_i\end{pmatrix}\\
\Sigma_{0\natural} &= \begin{pmatrix}\half\Gamma'_9 & 0\\ 0 & -\half
\Gamma'_9\end{pmatrix}\qquad & \Sigma_{09} &=
\begin{pmatrix}0 & -\half \Gamma'_9 \\ -\half \Gamma'_9 & 0
\end{pmatrix}\\
\Sigma_{9\natural} & = \begin{pmatrix}0 & -\half \1 \\ \half \1 & 0
\end{pmatrix}~,&&
\end{alignat*}
where $\Sigma'_{ij}$ are the generators of $\so(8)\subset\Cl(8)$.  In
particular, notice that as mentioned above, the representation
$\Delta$ breaks up under $\Spin(8)$ as two copies each of the spinor
representations.  In more traditional notation, under the embedding
$\Spin(10,1) \supset \Spin(8)$, we have
\begin{equation}\label{eq:spin8branching}
\repre{32} = 2\,\repre{8}_{\mathbf{s}}\oplus
2\,\repre{8}_{\mathbf{c}}~.
\end{equation}

\subsection{The spinor isotropy group $G$}

In this section we describe the non-reductive Lie subgroup $G \subset
\Spin(10,1)$ in equation \eqref{eq:exotic} which leaves a spinor
invariant.  We will exhibit its Lie algebra (and hence the Lie group
itself) inside the Clifford algebra $\Cl(1,10)$ constructed above.

Consider a spinor $\varepsilon$ of the form
\begin{equation}\label{eq:spinor}
\varepsilon = \begin{pmatrix}
              \psi_+ \\ 0
              \end{pmatrix}~,
\end{equation}
where $\psi_+$ is a positive chirality spinor of $\Cl(8)$; that is,
$\Gamma'_9\psi_+ = \psi_+$.  The vector $v^A = \Bar\varepsilon
\Gamma^A \varepsilon$ obeys $v^i =v^9 = v^+ = 0$ and $v^- = - 2
\|\psi_+\|^2$, where $\|\psi_+\|^2 = \Bar\psi_+ \psi_+$.  Therefore
the vector $v$ and hence the spinor $\varepsilon$ are null and by the
discussion in Section~\ref{sec:holonomy} the spinor isotropy should be
the group $G$ in equation \eqref{eq:exotic}.  Let us prove this.  It
is easy to compute the isotropy subalgebra $\fg \subset \so(10,1)$ of
$\varepsilon$ from the explicit form of the generators of $\so(10,1)$
given above.  After a little bit of algebra we obtain that the most
general element of $\fg$ is given by
\begin{equation}\label{eq:isotropy}
\half a_{ij} \Sigma_{ij} + b_\mu \Sigma_{-\mu}~,
\end{equation}
where $\Gamma_- = \Gamma_0 - \Gamma_\natural$, $b_\mu$
($\mu=1,\dots,9$) are arbitrary and $a_{ij}$ ($i,j=1,\dots,8$) are
such that the $\so(8)$ element $\half a_{ij} \Sigma'_{ij}$ actually
belongs to the isotropy subalgebra of $\psi_+$.  Because $\Spin(8)$
acts transitively on the unit sphere in both its spinor (as well as
the vector) representations, the isotropy subalgebras of every spinor
are conjugate, hence isomorphic.  This implies that the isotropy
subgroup of $\psi_+$ is a $\Spin(7)$ subgroup: one which decomposes
the positive chirality spinor representation of $\Spin(8)$ but keeps
the vector and the negative chirality representations irreducible.  We
call this subgroup $\Spin^+(7)$ to distinguish it from the other two
(conjugacy classes of) $\Spin(7)$ subgroups of $\Spin(8)$: $\Spin(7)$,
which leaves the spinor representations irreducible but splits the
vector representation, and $\Spin^-(7)$ which splits the negative
chirality spinor representation but leaves the positive chirality and
the vector representations irreducible.  This means that $\half
a_{ij}\Sigma_{ij}$ belong to the $\so^+(7)$ subalgebra of $\so(8)$.
Computing the Lie bracket of elements of the form \eqref{eq:isotropy},
and using that $\Gamma_-$ squares to zero, we obtain the following
structure
\begin{equation*}
\fg \cong \left ( \so^+(7) \ltimes \RR^8 \right) \times \RR~,
\end{equation*}
where $\RR^8$ is abelian and $\so^+(7)$ acts on it as a spinor.
Notice that $\RR$ is in the centre, and that $\RR^8$ is an abelian
ideal, whence this Lie algebra is not reductive.  Exponentiating
inside $\Cl(1,10)$ we obtain the simply-connected 30-dimensional
non-reductive Lie subgroup $G \subset\Spin(10,1)$ given in equation
\eqref{eq:exotic}:
\begin{equation}
  \label{eq:exotic2}
  G \cong \left( \Spin^+(7) \ltimes \RR^8 \right) \times \RR~.
\end{equation}

\subsection{The group $G$ as Lorentz transformations}

To gain some intuition about the group $G$, we now investigate the
action of $G$ on Minkowski spacetime $\MM^{11}$.  We will see that it
is generated by rotations and null rotations and that it acts
reducibly yet indecomposably.  It is convenient to parametrise $G$,
which as a manifold is diffeomorphic to $\Spin(7) \times \RR^9$, as
follows:
\begin{equation}
  \label{eq:Gparam}
  \Cl(1,10) \supset G \ni g = \exp\left(c_\mu \Sigma_{-\mu}\right)\,
  \sigma~,
\end{equation}
where $\sigma\in\Spin^+(7)$ and $\mu$ runs from $1$ to $9$.  Notice
that the exponential only consists of two terms because $\Gamma_-^2 =
0$:
\begin{equation*}
\exp\left( c_\mu \Sigma_{-\mu}\right) = 1 + c_\mu \Sigma_{-\mu} =
1 + \half c_\mu \Gamma_\mu\Gamma_-~.
\end{equation*}
The composition of group elements follows the standard semidirect
product structure:
\begin{equation*}
\exp\left( c_\mu \Sigma_{-\mu}\right)\,\sigma\,\exp\left( d_\mu
\Sigma_{-\mu}\right)\,\tau = \exp\left( \left(c_\mu + \sigma\cdot
d_\mu\right) \Sigma_{-\mu}\right)\,\sigma\tau~,
\end{equation*}
where $\sigma,\tau\in\Spin^+(7)$ and $c_\mu, d_\mu\in\RR^9$.

Let $g\in G$ be as above.  Its action on the basis $\{e_A\}$ is given
by
\begin{align*}
g\cdot e_i &= \sigma\cdot e_i - \sigma^{-1} \cdot c_i\,e_-\\
g\cdot e_9 &= e_9 - c_9 e_-\\
g\cdot e_\natural &= e_\natural - c_\mu e_\mu + \half c^2 e_-\\
g\cdot e_0 &= e_0 - c_\mu e_\mu + \half c^2 e_-~,
\end{align*}
where $c^2 = c_\mu c_\mu$.  Notice that for nonzero $c_\mu$ exactly
one null direction ($e_-$) is left invariant, so that the
transformation is a null rotation.  From these formulae above one can
determine the space of $G$-invariant forms.  The results are
summarised in Table~\ref{tab:forms}, where $\{e_A^*\}$ is a canonical
dual basis to $\{e_A\}$, $\Omega$ is the Cayley form (the self-dual
$\Spin^+(7)$-invariant $4$-form), $\vol_8 = e_1^* \wedge \dots \wedge
e_8^*$ and $\vol_{11}= e_0^* \wedge e_1^* \wedge \dots \wedge e_9^*
\wedge e_\natural^*$ are the eight- and eleven-dimensional volume
forms, respectively.

\begin{table}[h!]
\centering
\setlength{\extrarowheight}{5pt}
\begin{tabular}{|c|>{$}l<{$}|}
\hline
Degree& \multicolumn{1}{c|}{Invariant forms}\\
\hline\hline
0 & 1\\
1 & e_-^*\\
2 & e_-^*\wedge e_9^*\\
5 & e_-^*\wedge\Omega\\
6 & e_-^*\wedge e_9^*\wedge\Omega\\
9 & e_-^*\wedge \vol_8\\
10& e_-^*\wedge e_9^* \wedge \vol_8\\
11& \vol_{11}\\[5pt] \hline
\end{tabular}
\vspace{8pt}
\caption{$G$-invariant forms.}
\label{tab:forms}
\end{table}

Decomposing the spinor representation $\Delta$ under $\Spin^+(7)$, there
are precisely two linearly independent $\Spin^+(7)$-invariant spinors.
The null rotations in $\RR^9$ preserve only one of them.  Therefore
$G$ leaves invariant exactly one spinor (up to scale)---the spinor
$\varepsilon$ in \eqref{eq:spinor}, where $\psi_+$ is the
$\Spin^+(7)$-invariant spinor in that representation.

\subsection{Some relevant subgroups of $G$}

Bryant \cite{Bryant-spinors} has shown that the Lie group $G$ in
equation \eqref{eq:exotic2} is a possible holonomy group for
indecomposable eleven-di\-men\-sion\-al lorentzian manifolds.  There
are other subgroups of $G$ which are also possible holonomy groups and
which still act indecomposably on $\MM^{11}$.  These subgroups $H
\subset G$ are of the form
\begin{equation}
  \label{eq:subexotic}
  \left( K \ltimes \RR^d \right) \times \RR^{9-d}~,
\end{equation}
where $K\subset \Spin^+(7) \cap \Spin(d)$ is a possible holonomy group 
of a $d$-dimensional riemannian manifold.  Such subgroups can be read
off from Table~\ref{tab:static}, for $d\leq 8$.

The Lie algebra $\fh$ of $H$, given by
\begin{equation*}
  \left( \fk \ltimes \RR^d \right) \times \RR^{9-d}~,
\end{equation*}
is embedded in $\so(10,1)$ according to equation \eqref{eq:isotropy}
but where now $\half a_{ij} \Sigma_{ij} \in \fk \subset \so^+(7) \cap
\so(d)$.

We choose to parametrise group elements $h\in H$ as in equation
\eqref{eq:Gparam}
\begin{equation*}
  h = \exp\left(c_\mu \Sigma_{-\mu}\right)\, \sigma~,
\end{equation*}
but where now $\sigma\in K$.  The action of $h\in H$ on $\MM^{11}$ is
given by
\begin{align*}
h\cdot e_i &= \sigma\cdot e_i - \sigma^{-1} \cdot c_i\,e_-\\
h\cdot e_m &= e_m - c_m e_-\\
h\cdot e_\natural &= e_\natural - c_\mu e_\mu + \half c^2 e_-\\
h\cdot e_0 &= e_0 - c_\mu e_\mu + \half c^2 e_-~,
\end{align*}
where $i=1,\dots,d$ and $m=d{+}1,\dots,9$.

We can compute the space of $K$-invariant forms as before.  We will
not be so explicit as to actually list the forms, but simply notice
that the space of $H$-invariant forms on $\MM^{11}$ is given in terms
of the space of $K$-invariant forms on $\RR^d$ by:
\begin{equation}
  \label{eq:Hforms}
  \left(\bigwedge \MM^{11\,*} \right)^H \cong \RR 1 \oplus \RR
  \vol_{11} \oplus  e^*_- \wedge \left( \bigwedge \VV \otimes \left(
  \bigwedge \RR^{d\,*} \right)^K \right)~,
\end{equation}
where $\VV$ is the subspace of $\left(\MM^{11}\right)^*$ spanned by
the $e^*_m$ for $m=d{+}1,\dots,9$.  For example, for $d=7$ and
$K=G_2$, the $G_2$-invariant forms are $1$, $\phi$, $\star\phi$ and
$\vol_7$, where $\phi$ is the associative $3$-form and $\star\phi$ is
its seven-dimensional dual: the co-associative $4$-form.  The
$H$-invariant forms are given in Table~\ref{tab:formsG2} below.  One
can write down similar tables for the other subgroups $H$, but we will 
not do so here.

\begin{table}[h!]
\centering
\setlength{\extrarowheight}{5pt}
\begin{tabular}{|c|>{$}l<{$}|}
\hline
Degree& \multicolumn{1}{c|}{Invariant forms}\\
\hline\hline
0 & 1\\
1 & e_-^*\\
2 & e_-^*\wedge e_m^*\\
3 & e_-^*\wedge e_8^* \wedge e_9^*\\
4 & e_-^*\wedge\phi\\
5 & e_-^* \wedge \phi \wedge e_m^*~,\quad e_-^* \wedge \star\phi\\
6 & e_-^* \wedge \phi \wedge e_8^* \wedge e_9^*~,\quad e_-^* \wedge
\star\phi \wedge e_m^*\\ 
7 & e_-^* \wedge \star\phi \wedge e_8^* \wedge e_9^* \\
8 & e_-^* \wedge \vol_7 \\
9 & e_-^* \wedge \vol_7 \wedge e_m^*\\
10& e_-^* \wedge \vol_7 \wedge e_8^* \wedge e_9^*\\
11& \vol_{11}\\[5pt] \hline
\end{tabular}
\vspace{8pt}
\caption{Forms on $\MM^{11}$ invariant under $H=\left(G_2 \ltimes
\RR^7\right) \times \RR^2$.  Here $m$ takes the values $8$ and
$9$.}
\label{tab:formsG2}
\end{table}

The action on the eleven-dimensional spinor representation $\Delta$
can be worked out from the explicit expression \eqref{eq:isotropy} as
before.  Decomposing $\Delta$ under the group $K$ we find $2 N_K$
parallel spinors, where $N_K =1$ for $K=\Spin^+(7)$, $N_K=2$ for
$K=\SU(4)$ and $G_2$, $N_K=3$ for $K=\Sp(2)$, $N_K=4$ for
$K=\Sp(1)\times\Sp(1)$ and $\SU(3)$, $N_K=8$ for $K=\Sp(1)$ and
$N_K=16$ for $K=\{1\}$.  The null rotations in $H$ preserve precisely
one half of these spinors, whence $H$ leaves invariant precisely $N_K$
spinors. This gives rise to the supersymmetry fractions $\nu \equiv
\frac1{32} N_K$ in Table~\ref{tab:mwaves}.

%

\begin{thebibliography}{10}

\bibitem{AFSS}
BS~Acharya, JM~Figueroa-O'Farrill, B~Spence, and S~Stanciu, \emph{Planes,
  branes and automorphisms: {II.} {B}ranes in motion}, JHEP \textbf{07} (1998),
  005, {\tt hep-th/9805176}.

\bibitem{BBI}
L~Bérard Bergery and A~Ikemakhen, \emph{On the holonomy of lorentzian
  manifolds}, Proc. Symp. Pure Math. \textbf{54} (1993), 27--40.

\bibitem{BKO}
EA~Bergshoeff, R~Kallosh, and T~Ortín, \emph{Supersymmetric string waves},
  Phys. Rev. \textbf{D47} (1993), 5444--5452, \texttt{hep-th/9212030}.

\bibitem{Besse}
AL~Besse, \emph{Einstein manifolds}, Springer-Verlag, 1987.

\bibitem{Brinkmann2}
HW~Brinkmann, \emph{Einstein spaces which are mapped conformally on each
  other}, Math. Ann. \textbf{94} (1925), 119--145.

\bibitem{Bryant-spinors}
RL~Bryant, \emph{Remarks on spinors in low dimension}, Unpublished notes, 1998.

\bibitem{CJS}
E~Cremmer, B~Julia, and J~Scherk, \emph{Supergravity in eleven dimensions},
  Phys. Lett. \textbf{76B} (1978), 409--412.

\bibitem{DNP}
MJ~Duff, BEW Nilsson, and CN~Pope, \emph{Kaluza--{K}lein supergravity}, Phys.
  Rep. \textbf{130} (1986), 1--142.

\bibitem{GSR}
C~Gabriel, P~Spindel, and M~Rooman, \emph{Chiral supersymmetric $pp$-wave
  solutions of {IIA} supergravity}, Phys. Lett. \textbf{B415} (1997), 54--62,
  \texttt{hep-th/9709142}.

\bibitem{HoTs}
GT~Horowitz and AA~Tseytlin, \emph{A new class of exact solutions in string
  theory}, Phys. Rev. \textbf{D51} (1995), 2896--2917, \texttt{hep-th/9409021}.

\bibitem{Mwave}
CM~Hull, \emph{Exact $pp$ wave solutions of eleven-dimensional supergravity},
  Phys. Lett. \textbf{139B} (1984), 39--41.

\bibitem{Hull-planewaves}
\bysame, \emph{Killing spinors and exact plane wave solutions of extended
  supergravity}, Phys. Rev. \textbf{D30} (1984), 334--338.

\bibitem{JuliaNicolai}
B~Julia and H~Nicolai, \emph{Null {K}illing vector dimensional reduction and
  galilean geometrodynamics}, Nucl. Phys. \textbf{B439} (1995), 291.

\bibitem{KSMH}
D~Kramer, D~Stephani, MAH MacCallum, and E~Herlt, \emph{Exact solutions of
  {E}instein's field equations}, Cambridge University Press, 1980.

\bibitem{LM}
HB~Lawson and ML~Michelsohn, \emph{Spin geometry}, Princeton University Press,
  1989.

\bibitem{Lidsey}
JE~Lidsey, \emph{The embedding of superstring backgrounds in {E}instein
  gravity}, Phys. Lett. \textbf{B417} (1998), 33.

\bibitem{Nahm}
W~Nahm, \emph{Supersymmetries and their representations}, Nucl. Phys.
  \textbf{B135} (1978), 149--166.

\bibitem{NW}
C~Nappi and E~Witten, \emph{A {WZW} model based on a non-semi-simple group},
  Phys. Rev. Lett. \textbf{71} (1993), 3751--3753, {\tt hep-th/9310112}.

\bibitem{dR}
G~De Rham, \emph{Sur la réducibilité d'un espace de {R}iemann}, Comm. Math.
  Helv. \textbf{26} (1952), 328--344.

\bibitem{Arkady}
AA~Tseytlin, \emph{String vacuum backgrounds with covariantly constant null
  {K}illing vector and 2-d quantum gravity}, Nucl. Phys. \textbf{B390} (1993),
  153--172, \texttt{hep-th/9209023}.

\bibitem{Urrutia}
LF~Urrutia, \emph{A new exact solution of classical supergravity}, Phys. Lett.
  \textbf{102B} (1981), 393--396.

\bibitem{Wu}
H~Wu, \emph{On the de~{R}ham decomposition theorem}, Illinois J. Math.
  \textbf{8} (1964), 291--311.

\end{thebibliography}
%

\providecommand{\bysame}{\leavevmode\hbox to3em{\hrulefill}\thinspace}

\end{document}